\documentclass[twocolumn]{aastex631}

\begin{document}

\title{The radio halo in PLCKESZ G171.94 $-$ 40.65: Beacon of merging activity}

\author[0009-0002-0373-570X]{R. Santra}
\affiliation{National Centre for Radio Astrophysics, Tata Institute of Fundamental Research, Pune 411007, India}

\affiliation{INAF - IRA, Via Gobetti 101, I-40129 Bologna, Italy; IRA - INAF, via P. Gobetti 101, I-40129 Bologna, Italy}

\author[0000-0003-1449-3718]{R. Kale}
\affiliation{National Centre for Radio Astrophysics, Tata Institute of Fundamental Research, Pune 411007, India}

\author[0000-0002-1634-9886]{S. Giacintucci}
\affiliation{Naval Research Laboratory, 4555 Overlook Avenue SW, Code 7213, Washington, DC 20375, USA}

\author[0000-0001-9110-2245]{D. R. Wik}
\affiliation{Department of Physics \& Astronomy, University of Utah, 115 South 1400 East, Salt Lake City, UT 84112, USA}

\author[0000-0002-8476-6307]{T. Venturi}
\affiliation{INAF - IRA, Via Gobetti 101, I-40129 Bologna, Italy; IRA - INAF, via P. Gobetti 101, I-40129 Bologna, Italy}

\author[0000-0003-1246-6492]{D. Dallacasa}
\affiliation{Dipartimento di Fisica e Astronomia, Università di Bologna, via P. Gobetti 93/2, 40129, Bologna, Italy}
\affiliation{INAF - IRA, Via Gobetti 101, I-40129 Bologna, Italy; IRA - INAF, via P. Gobetti 101, I-40129 Bologna, Italy}

\author[0000-0003-4046-0637]{R. Cassano}
\affiliation{INAF - IRA, Via Gobetti 101, I-40129 Bologna, Italy; IRA - INAF, via P. Gobetti 101, I-40129 Bologna, Italy}

\author[0000-0003-4195-8613]{G. Brunetti}
\affiliation{INAF - IRA, Via Gobetti 101, I-40129 Bologna, Italy; IRA - INAF, via P. Gobetti 101, I-40129 Bologna, Italy}

\author{D. C. Joshi}
\affiliation{National Centre for Radio Astrophysics, Tata Institute of Fundamental Research, Pune 411007, India}




\begin{abstract}

We present the first multi-frequency analysis of the candidate ultra-steep spectrum radio halo in the galaxy cluster PLCKESZ G171.94$-$40.65, using the upgraded Giant Metrewave Radio telescope (uGMRT; 400 MHz), and Karl G. Jansky Very Large Array (JVLA; 1-2 GHz) observations. Our radio data have been complemented with archival \textit{Chandra} X-ray observations to provide a crucial insight into the complex intracluster medium (ICM) physics, happening at large scales. We detect the radio halo emission to the extent of $\sim$ 1.5 Mpc at 400 MHz, significantly larger than previously reported, along with five tailed galaxies in the central region. We also report the discovery of an unknown diffuse source `U', at the cluster periphery, with an extent of 300 kpc. Using the available observations, we have found that the radio spectrum of the halo is well-fitted with a single power law, having a spectral index of $-1.36 \pm 0.05$, indicating that it is not an ultra-steep spectrum radio halo. Our low-resolution (25$''$) resolved spectral map shows an overall uniform spectral index, with some patches of fluctuations. The X-ray and radio surface brightness are morphologically co-spatial, with a slight extension along the northwest-southeast direction, seen in both maps. The radio and X-ray surface brightness indicates strong positive correlations, with sub-linear correlation slopes ($\sim$ 0.71). Multiple tailed galaxies and the radio halo indicate a high dynamical activity at the cluster central region.

\end{abstract}




\section{Introduction} \label{sec:intro}

Galaxy clusters contain non-thermal components (relativistic particles and magnetic fields) in the state of ionized (n$_{e}$ $\sim$ 10$^{-3}$ $-$ 10$^{-4}$ cm$^{-3}$) plasma mixed with the hot (10$^{7}$ $-$ 10$^{8}$ K) thermal Intracluster medium (ICM) \citep{RevModPhys.58.1}. In many dynamically disturbed clusters, the non-thermal plasma shows highly extended and diffuse synchrotron sources, broadly classified as radio relics and radio halos \citep[for a review;][]{2019SSRv..215...16V}. The non-thermal emission is not associated with any individual galaxy, suggesting a large-scale (in-situ) acceleration mechanism feeding the radio-emitting electrons \citep[for a review;][]{2014IJMPD..2330007B}. These radio sources have a very steep spectrum\footnote{S$_{\nu}$ $\propto$  $\nu ^{\alpha}$, where S$_{\nu}$ is the flux density and $\alpha$ for spectral index} ($\alpha$ \textless -1). Both halos and relics are observed mostly in highly disturbed clusters \citep[e.g.,][]{2001ApJ...553L..15B, 2008A&A...486..347G, 2010ApJ...721L..82C, 2011PhDT........14V}, suggesting that cluster dynamics may play a major role behind their occurrence. Despite the several efforts that have been made to understand radio halos and relics using both theoretical and cutting-edge observational methods, the exact formation mechanism is not completely understood \citep[e.g.,][]{2014IJMPD..2330007B}.

The current reference model for the formation of a radio halo indicates the re-energization of the pre-existing seed relativistic electrons, due to merger-driven turbulence \citep[i.e., re-acceleration models:][]{2001MNRAS.320..365B,petrosian2001nonthermal,2006AN....327..557C, 2007MNRAS.378..245B, 2013ApJ...771..131B, 2016MNRAS.458.2584B}. The overall signatures obtained via observations match quite well with the predictions of the turbulent re-acceleration model \citep{2021A&A...647A..51C,2023A&A...672A..43C}. On the other hand, the contribution from secondary electrons formed as a by-product of hadronic collisions \citep[e.g.,][]{1999NuPhS..70..495B} is constrained to be sub-dominant based on $\gamma$-ray upper-limits and the energetics of cosmic rays (CRs) \citep[e.g.,][]{2008Natur.455..944B,2012MNRAS.426...40B, 2017AIPC.1792b0009B, 2021A&A...650A..44B, 2024arXiv240509384O}. However, the re-acceleration of the secondary electrons is still a viable scenario \citep[e.g.,][]{2011MNRAS.410..127B, 2021A&A...648A..60A}.

Radio halos have been reported to have a very broad range of spectral indices, spanning from -1.0 to -2.1 \citep[for reviews;][]{2012A&ARv..20...54F, 2019SSRv..215...16V}. In particular, a class of radio halos with a very steep spectral index (\textless -1.5), known as Ultra-Steep Spectrum Radio Halos (USSRHs), are predicted by turbulent re-acceleration mechanism via less energetic mergers \citep[i.e., involving clusters with masses \textless 10$^{15}$~M$_{\odot}$; see, e.g.,][]{2006AN....327..557C, 2010ApJ...721L..82C}. The prototype example is Abell 521 \citep{2008Natur.455..944B, 2009ApJ...699.1288D, 2013A&A...551A.141M, 2024ApJ...962...40S}. The pure hadronic origin for explaining the occurrence of steep spectrum radio halos has difficulties, due to the requirement of a large energy budget from relativistic protons \citep[e.g.,][]{2008Natur.455..944B, 2021A&A...650A..44B}. The number of confirmed steep spectrum radio halos with a well-constrained spectrum of \textgreater 3 data points has increased due to sensitive observations from hundreds of MHz to a few GHz \citep[e.g.,][]{2013A&A...551A.141M, 2018MNRAS.473.3536W, 2021A&A...650A..44B, 2023A&A...669A...1R, 2023A&A...678A.133B}. Other candidate USSRHs still lack confirmation of their spectral index \citep[e.g.,][]{2011A&A...534A..57G, 10.1111/j.1365-2966.2012.21570.x, 2019JApA...40...17S, 2021MNRAS.505..480R} due to the unavailability of sensitive multi-frequency observations. Therefore, it is important to affirm the candidates and increase the population of these USSRHs, with robust estimation of their spectral index, to understand the turbulent re-acceleration in detail. Several dedicated searches for the cluster diffuse sources have taken place over the last decade, and the number of radio halos and relics has increased rapidly \citep[e.g.,][]{2013A&A...557A..99K, 2015A&A...579A..92K, 2021PASA...38...31D,2021NatAs...5..268D, 2022A&A...660A..78B, 2022A&A...657A..56K, 2022MNRAS.514.5969K, 2024PASA...41...26D}. However, due to their low surface brightness, most of them do not have detailed spectral information, which limits our understanding of the complex chain of the physical process that bridges the cluster merger and the formation and evolution of turbulence at local ICM scales.

The cluster PLCKESZ G171.94$-$40.65 (hereafter PLCK171), located at a redshift of 0.27, was discovered as a part of the Planck Early Sunyaev-Zel'dovich (ESZ) all-sky survey \citep{2011A&A...536A...8P}. The clusters in this survey were detected by observing the distortion in the cosmic microwave background (CMB) spectrum due to scattering with hot ICM electrons, known as the Sunyaev-Zel’dovich (SZ) effect \citep{1972CoASP...4..173S}. The detection of the cluster was later confirmed using the \textit{XMM-Newton} X-ray observations \citep{2011A&A...536A...9P}. The mass (M$_{500}$) of the cluster is 1.05$_{-1.3}^{+1.0}$ $\times$ 10$^{15}$ M$_{\odot}$, with a very high X-ray luminosity of L$_{\rm X [0.1 - 2.4~\rm keV]}$ = 1.13 $\times$ 10$^{45}$ erg s$^{-1}$, and a temperature of 11.0$_{-0.5}^{+0.9}$ keV \citep{2017ApJ...841...71G}.

\begin{table}
\centering
  \caption{Summary of the observations.}
  \begin{tabular}{@{}cccc@{}}
    \hline\hline
     Telescope &  Frequency & Bandwidth   & Obs. time  \\
    \hline
 
  uGMRT  &   400 MHz & 200 MHz  & 8 hr.  \\

legacy GMRT &  610 MHz & 32 MHz & 8 hr.   \\

 JVLA C-array  &   1.4 GHz &  1 GHz   & 2 hr.  \\

 JVLA D-array  &   1.4 GHz & 1 GHz & 2 hr.  \\

 \hline

  \end{tabular}

\tablecomments{The legacy GMRT observations were taken in 235/610 MHz simultaneous mode.}
  
  \label{obs_table}
  
\end{table}

\citet{2013ApJ...766...18G} reported the discovery of a giant ($\sim$ 1 Mpc) radio halo with the legacy GMRT 235 MHz and NVSS (NRAO VLA Sky Survey) data at 1400 MHz. They estimated an integrated spectral index of the halo of $\sim$ -1.84 $\pm$ 0.14, which is very steep compared to the observed spectral index of a typical radio halo. However, the spectral index calculation of the PLCK171 halo was affected by several factors, including the short usable time of the 235 MHz observation ($\sim$ 1 hr) and a short integration time of the NVSS pointing. Also, the poor resolution of the NVSS prevents an accurate subtraction of the radio galaxies embedded in the halo emission. Targeted observations over a wide frequency range, with higher angular resolution and deeper exposures, are essential before any conclusive estimation of the spectral index of this halo can be made.

Here, we improve the multi-frequency radio coverage of PLCK171 and compare it to existing X-ray data. We present a new upgraded Giant Meterwave Radio Telescope (uGMRT) at 400 MHz and Karl G. Jansky Very Large Array (JVLA) data in L-band (1$-$2 GHz). The 235, 325, and 610 MHz analysis is presented in \citet{Joshi_21}. In addition to the radio data, we present archival \textit{Chandra} X-ray observations to understand the dynamics of the ICM along with studying correlations between the thermal and non-thermal emission of the cluster.

The paper is organized as follows: In Sec.~\ref{sec:2} we provide a brief overview of the observations and data analysis procedures to obtain sensitive images. In Sec.~\ref{sec:3}, we discuss structures in the radio images and the results obtained from spectral analysis. The radio vs.\ X-ray correlations are described in Sec.~\ref{sec:4}. Then we discuss our results in the context of the mergers happening in this cluster in Sec.~\ref{sec:5} and summarise the findings in Sec.~\ref{sec:6}. Throughout this paper, we have adopted a flat $\Lambda$CDM cosmology with H$_{0}$ = 70 km s$^{-1}$, $\Omega_{m}$ = 0.3, $\Omega_{\Lambda} = 0.7$. At the redshift of PLCK171, 1$''$ corresponds to a linear scale of 4.13 kpc.

\begin{table}
\centering
  \caption{Summary of the radio images.}
  \begin{tabular}{@{}ccccc@{}}
    \hline\hline
  Freq & Name & Beam & Robust  & rms   \\

  (MHz) &  & $''$ &  & ($\mu$Jy~beam$^{-1}$) \\ 
  \hline
400 & IMG1 & 9$''$ $\times$ 8$''$  & 0.5& 29.5   \\

 & IMG2 & 21$''$ $\times$ 21$''$ & -0.5 & 50.0  \\

 \hline
 
 610  & IMG3 & 5$''$ $\times$ 4$''$ & 0.5& 37.0  \\ 

\hline
1400 & IMG4 & 21$''$ $\times$ 21$''$  & 0.5 & 23.0   \\

     & IMG5 & 35$''$ $\times$ 35$''$  & 0.5 & 29.0   \\

 \hline
  \end{tabular}
  \tablecomments{The IMG2 is made considering a inner \textit{uv}-cut of 0.2k$\lambda$, and \textit{uv}-taper of 20$''$.}
  \label{image_summary}
\end{table}

\section{Observations and data reduction} \label{sec:2}

PLCK171 was observed with the legacy GMRT (observation code: 23\_014), the upgraded GMRT (observation code: 43\_045), and the JVLA (proposal code: 13A-152) over a broad frequency range (Table ~\ref{obs_table}). We used real-time radio frequency interference (RFI) filtering for the uGMRT observation, developed for mitigating the broadband RFI \citep{2023JApA...44...37B,2022JAI....1150008B, 2024ApJ...962...40S}.

\subsection{uGMRT and legacy GMRT}

We have re-processed only the legacy GMRT data at 610 MHz using  \texttt{CAPTURE\footnote{\url{https://github.com/ruta-k/CAPTURE-CASA6}}}, a \texttt{CASA} based automated pipeline for the GMRT data reduction \citep{2021ExA....51...95K}. After following the standard calibration routines (we have adopted the \citealt{2017ApJS..230....7P} fluxscale for the primary calibrator), the calibration was applied to the target field, and after splitting, it was further flagged using the automated flagging tools available in CASA. Some low-level RFI was cleaned further using the aoflagger \citep{2010ascl.soft10017O}. The channels were averaged to reduce the data volume but still prevent bandwidth smearing. The target visibilities were then imaged using the CASA task \texttt{tclean} with wide-field imaging algorithms.  

Initially, the uGMRT observations were processed using the \texttt{CAPTURE}, however, the observations were heavily affected by the phase residual errors, created by two very bright sources (18$'$, 30$'$ away from the phase center) in the field of view, requiring direction-dependent calibration to mitigate those. Therefore, wideband GMRT data was processed using the Source Peeling and Atmospheric Modeling (\texttt{SPAM}; \citealt{2017A&A...598A..78I} ) pipeline\footnote{\url{http://www.intema.nl/doku.php?id=huibintemaspampipeline}}. The wide-band data were split into six narrow-band (with $\sim$ 32 MHz bandwidth), which were processed independently using the \texttt{SPAM} pipeline. The flux density scale of the primary calibrator 3C48 was set according to \cite{2012MNRAS.423L..30S}. After flux density calibration, the data were flagged, and corrected for the bandpass profile. For the correction of the phases for the target field, we started with a sky model obtained from the GMRT narrowband observation and estimated the direction-dependent gains for each of the bright sources. For further details on the data analysis, please refer to \citet{2021A&A...654A..41R}. The final deconvolution was performed in \texttt{CASA} using \texttt{nterms}=2, and \texttt{wprojection} along with a \texttt{robust= 0.5} (Table \ref{image_summary}), to better highlight the extended emission.

\begin{figure*}
    \centering
    \includegraphics[width=0.90\textwidth]{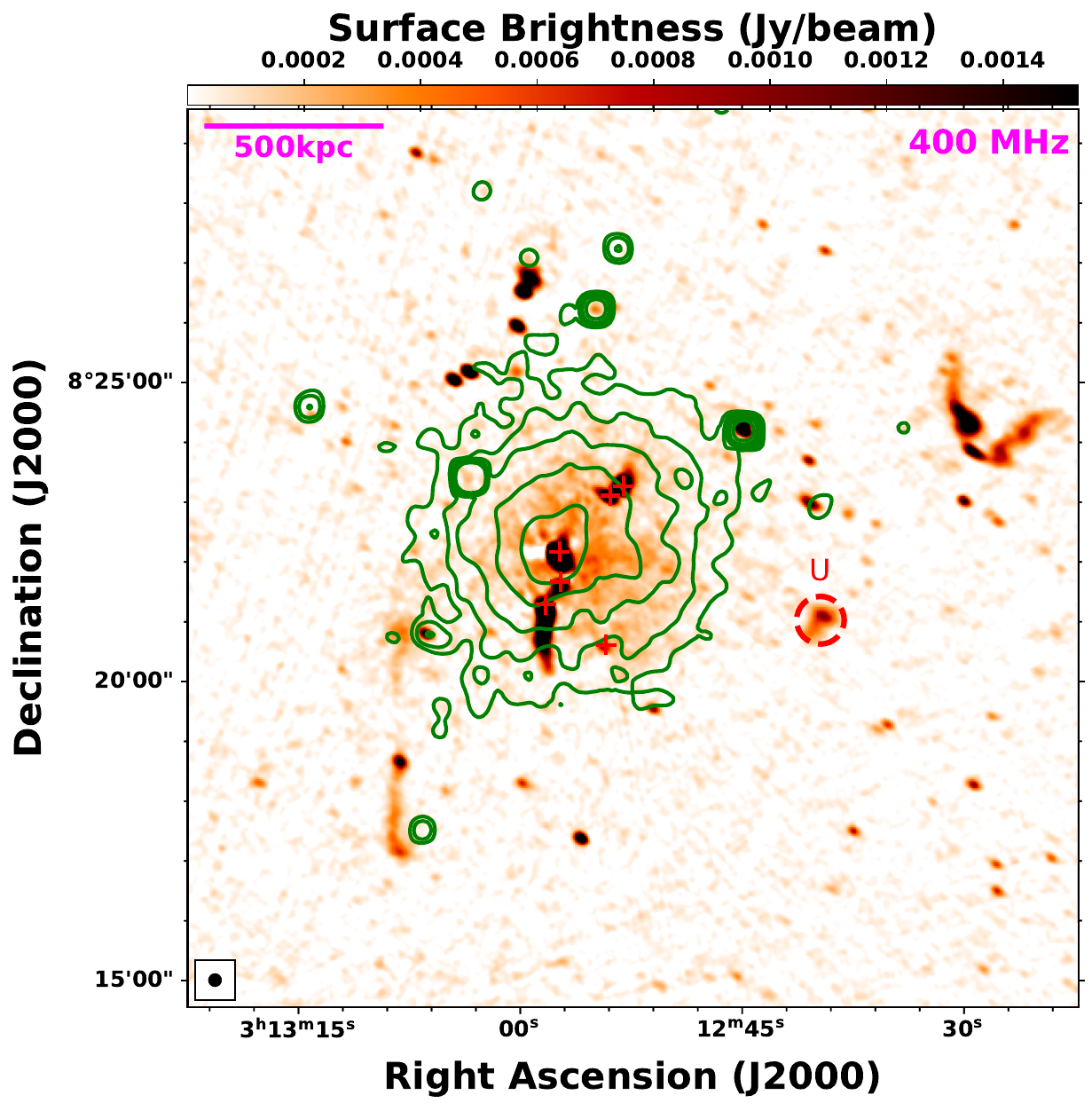}
    \caption{uGMRT 400 MHz full resolution image (color) IMG1 obtained with \texttt{Robust=0.5} is shown here. We have marked the tailed radio galaxies and discrete sources with red `+' signs. We highlight the newly detected diffuse source, named `U', with a dashed red circle. In the central region, low surface brightness radio halo emission is seen. The green contours show the cluster x-ray emission as imaged by \textit{Chandra} (Figure~\ref{img:8a})}.
    
    \label{img:1}
\end{figure*}

\begin{figure*}

    \includegraphics[width=10cm, height =10cm]{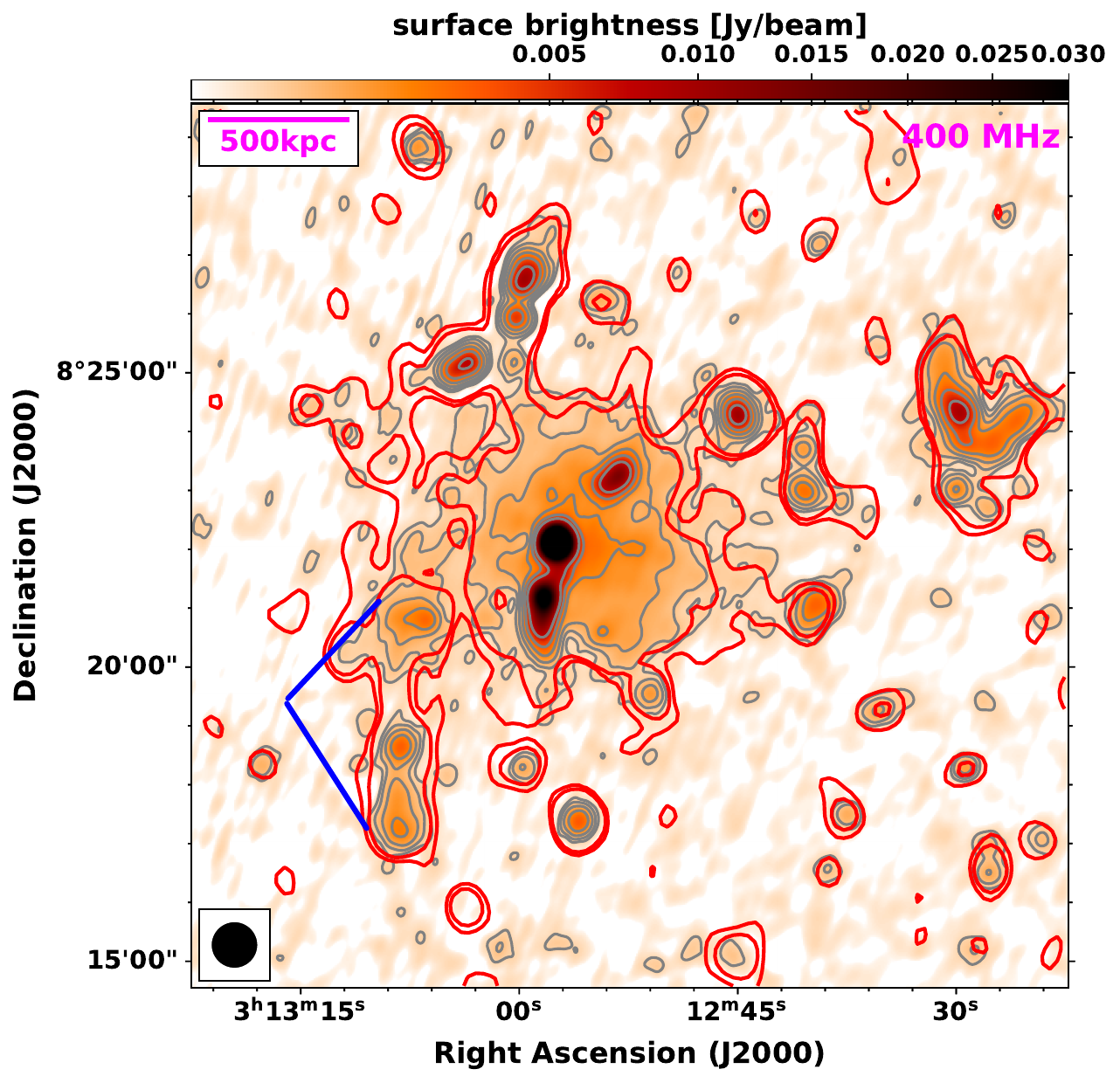}
    \includegraphics[width=8cm, height=10cm]{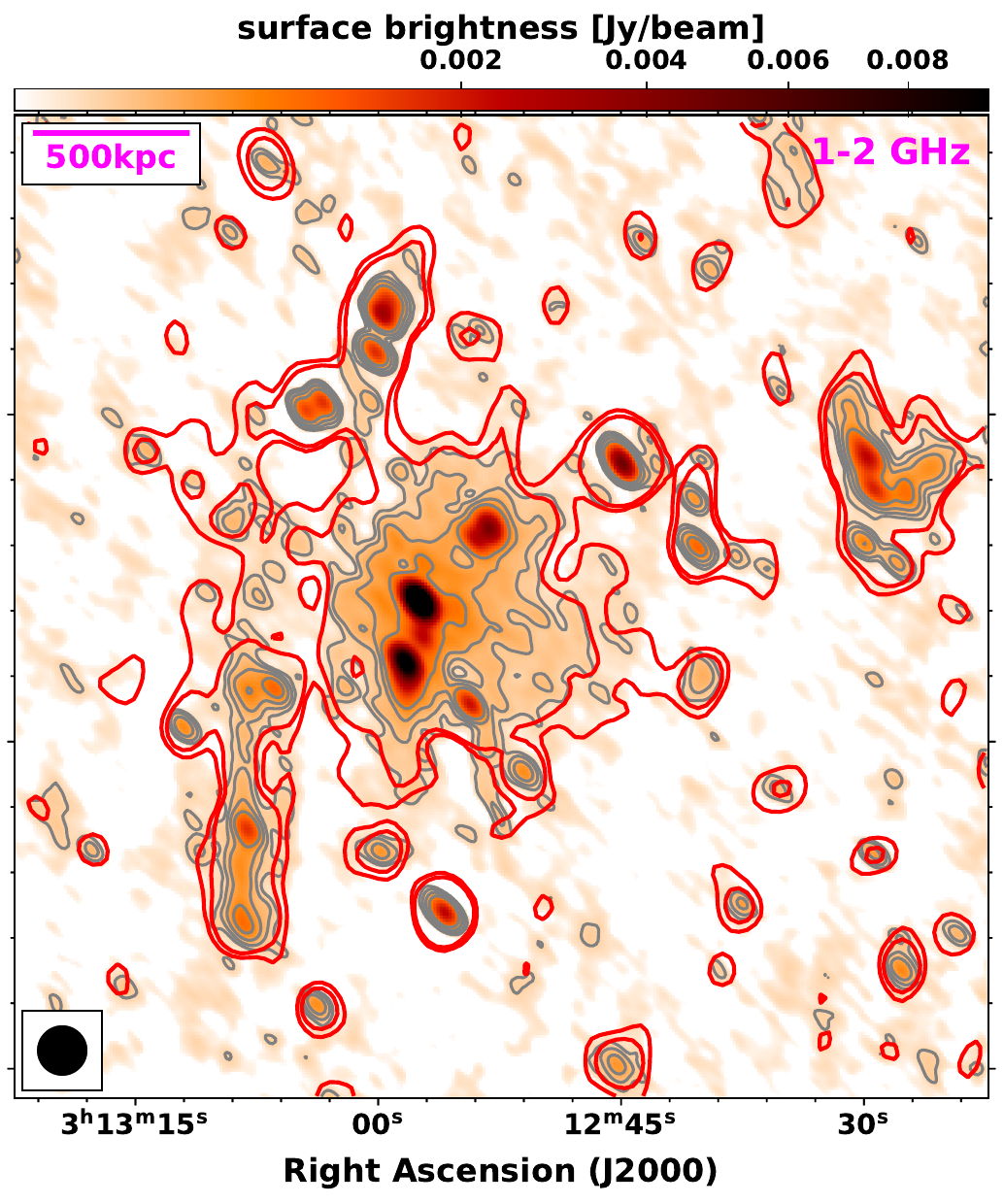}
     
    \caption{ \textit{Left:} The cluster central diffuse sources are shown in a low resolution (21$''$) image (IMG2) at 400 MHz. The Contour (grey) levels are drawn at [1, 2, 4, 8...]$\times$ 3$\sigma_{\rm rms}$, where $\sigma_{\rm rms}$ = 50 $\mu$Jybeam$^{-1}$ is the noise level. The overlaid red contour is from the JVLA D-array image (IMG5), drawn at a significance of [1, 2] $\times$ 3$\sigma_{\rm rms}$, with $\sigma_{\rm rms} = 29~\mu \rm Jy~\rm beam^{-1}$, with a beam size of 35$''\times 35''$. The blue-lined sector marks the AGN, which is seen to be connected with the radio halo. \textit{Right:} Here we show the 21$''$ JVLA C and D array combined image (IMG4) of the cluster field at 1400 MHz. The grey contours are drawn at a similar significance level, with a $\sigma_{\rm rms}$ = 23~$\mu$Jy~beam$^{-1}$. The red contours are drawn from JVLA D-array images, just for a comparison of the extension of the diffuse emission.}  
    \label{img:2a}
    \label{img:2b}
\end{figure*}

\subsection{VLA}

The data were recorded with the default wide-band setup of 16 spectral windows spanning the entire bandwidth, with each spectral window having 64 channels. We have processed the C and D arrays independently with the similar procedure mentioned below. The data reduction was carried out with the \texttt{CASA} v6.5 \citep{2007ASPC..376..127M} using 3C147 as the bandpass and flux calibrator (fluxscale adopted \citealt{2013ApJS..204...19P} fluxscale) and J1927+6117 as the phase calibrator. Briefly, we analyze each single spectral window (spw) dataset individually, and after the removal of radio frequency interference (RFI) using the \texttt{tfcrop} mode, we calibrate the antenna delays, and bandpass, and correct the antenna gains. Finally, we combined all the spectral windows and applied all the solutions to the target averaging in time and frequency (factor of two and four, respectively) for the reduction of the data volume. We performed self-calibration (3 rounds of phase-only) on each configuration to refine the amplitude and phase variation. For each of the imaging rounds during the self-calibration, we used \texttt{Briggs} weighting \citep{1995PhDT.......238B}, multi-frequency synthesis (MS-MFS) with \texttt{nterms=2} \citep{2011A&A...532A..71R}, and 256 W-projection planes \citep{2008ISTSP...2..647C}. Finally, we combined the self-calibrated visibilities for the C and D array configurations to produce the final deep images of the cluster, by performing a final self-calibration with a long solution interval to align the data sets. We have tabulated all the image properties at different frequencies, with their properties in Table~\ref{image_summary}.

\subsection{Removal of the point sources}

The diffuse radio emission in PLCK171 is difficult to measure since it embeds several bright radio galaxies. We have subtracted the point sources for the 400 and 1400 MHz observations. For the removal of the discrete sources, their models were created by applying a \textit{uv}-cut of \textgreater 4 k$\lambda$ (angular scale of 50$''$). The baseline-restricted model component corresponding to the ``discrete'' sources, was Fourier transformed and subtracted from the observed data using \texttt{uvsub} in \texttt{CASA}. After subtraction, the visibilities were imaged using \texttt{tclean} with a \textit{uv} baseline of \textless 10 k$\lambda$, \texttt{robust=0.5}, and \texttt{multi-scale} deconvolver to highlight the extended emission. We have followed a similar subtraction procedure for the uGMRT and JVLA. The consistency of the point source subtraction process is also verified by comparing the flux density of the radio halo with the value obtained by algebraically subtracting the flux densities of the embedded compact sources from the total (halo $+$ compact sources) flux density, finding a good agreement (6\% for uGMRT, and 8\% for the JVLA).

\subsection{Flux density scales}

The overall flux scale for all observations (uGMRT, and VLA) was checked by comparing the spectra of compact sources in the field of view. We assume the flux density uncertainty is $10\%$ for the GMRT data \citep{2017ApJ...846..111C}, and $2.5\%$ for the JVLA data. The uncertainty on the flux density measurement is defined by:
\begin{equation}
    \Delta \rm S = \sqrt{(\rm f \cdot \rm S)^{2} + (N_{\rm beam}(\sigma_{\rm rms})^{2}) + (\sigma_{\rm sub} \cdot \rm S)^{2}},
    \label{eq:1}
\end{equation}
where f is the absolute flux density calibration errors, $\rm S$ is the flux density, $\rm N_{\rm beam}$ is the number of beams, and $\sigma_{\rm rms}$ is the rms noise, and $\sigma_{\rm sub}$ is the point source subtraction error. We adopted $\sigma_{\rm sub}$ = 6\% and 8\% for the uGMRT and JVLA respectively. Note that, all the images are corrected for the primary beam using the task  \texttt{ugmrtpb}\footnote{\url{https://github.com/ruta-k/uGMRTprimarybeam-CASA6}} for all the GMRT images, and \texttt{widebandpbcor} for the JVLA images.

\subsection{Chandra}

PLCK171 was observed by {\em Chandra} during Cycle 14 using the ACIS-I detector (Obs ID 15302) in VFAINT mode for a total exposure time of 27 ks. We reprocessed the Level-1 ACIS event file from the {\em Chandra} archive using CIAO 4.13 and CALDB 4.9.5, following the Chandra analysis threads\footnote{\url{https://cxc.harvard.edu/ciao/threads/index.html}}. We checked the observation for periods of high background using the LC$_{-}$CLEAN script and found no significant problems. To model the detector and sky background, we used the blank-sky datasets from the CALDB appropriate for the date of the observation, reprojected, and normalized using the ratio of the observed to blank-sky count rates in the 9.5$-$12 keV band. We finally obtained an exposure-corrected and background-subtracted image of the cluster in the $0.5-4.0$ keV band. Point sources were identified using the CIAO \texttt{wavdetect} task and subtracted from the image, and the resulting holes were 
refilled with local background using the \texttt{dmfilth}.

\begin{figure}
	\includegraphics[width=\columnwidth]{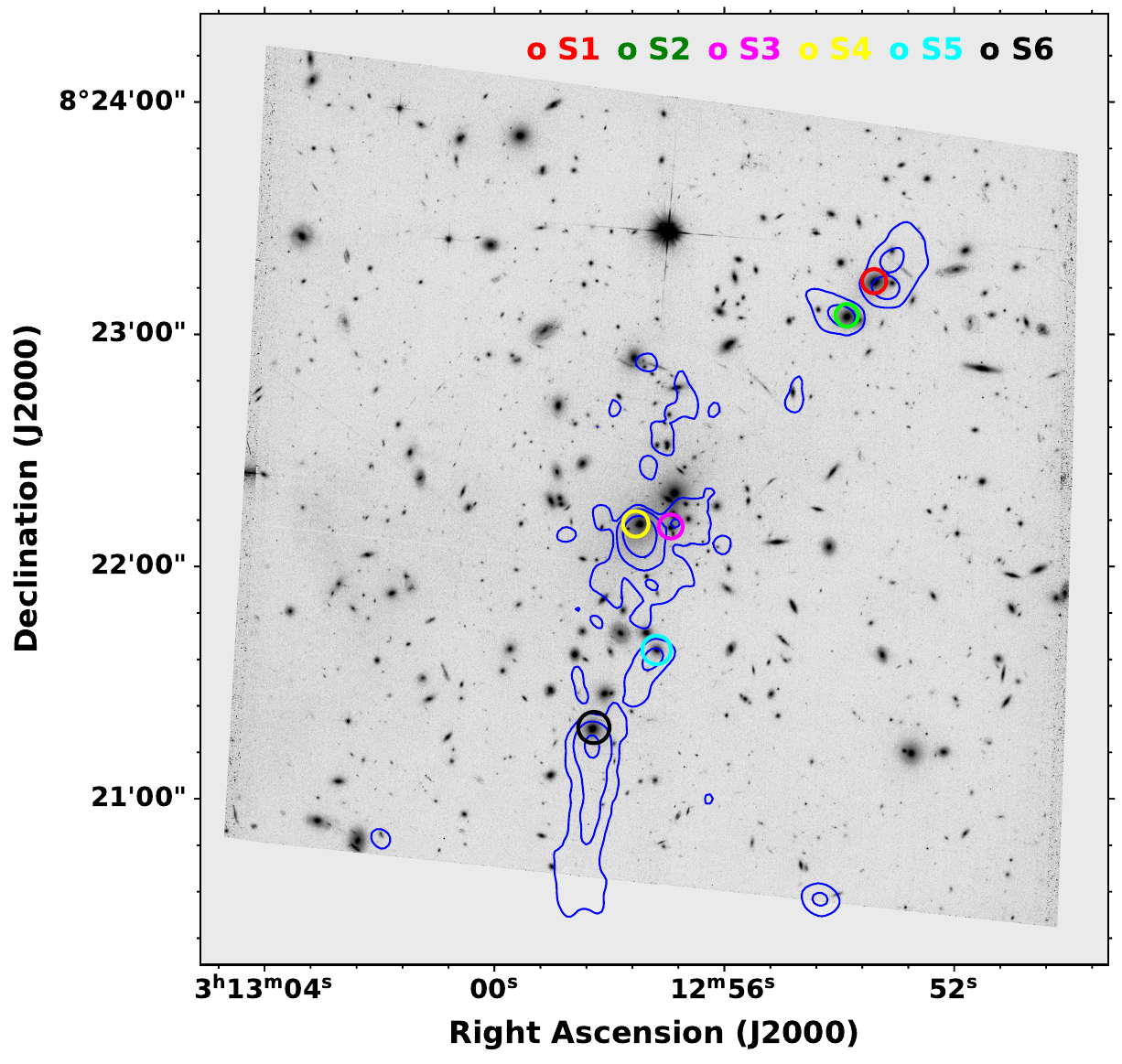}
        \caption{The grey scale image shows the PLCK171 core region, observed with HST/ACS filter \citep{2018ApJ...858...42A}. The blue contours are from the high-resolution image at 610 MHz (IMG3) and drawn at [3, 9, 27, 81]$\times$ $\sigma_{\rm rms}$, where $\sigma_{\rm rms} = 37~\mu$Jy~beam$^{-1}$ is the noise level, with a beam size of the image is 5.5$''\times 4.5''$. The different colored circle indicates the detection of the optical counterpart of each radio source. Their naming conventions (S7 is out of the HST image) are annotated in the image, and similar to  \citet{2013ApJ...766...18G}.} 
    \label{img:3}
\end{figure}

\section{Radio Results} \label{sec:3}

\begin{table*}
  \centering
  \caption{List of the compact sources and tailed galaxies }
  \begin{tabular}{@{}cccccccccc@{}}
    \hline\hline

& Source & RA & DEC  & S$_{400 \rm MHz}$ & S$_{610 \rm MHz}$ & S$_{1400 \rm MHz}$ & Spectral index & morphology \\
&  & hh mm ss & $^{\circ}$ $'$ $''$ & (mJy) & (mJy) & (mJy) &   \\
    \hline
& S1 & 03 12 53.2  & +08 23 12  & 18.72 $\pm$ 1.60 & 14.50 $\pm$ 2.20 & 5.89 $\pm$ 0.85 & -0.78 $\pm$ 0.09 & NAT\\ 
 
  & S2 & 03 12 53.9  & +08 23 05   & 7.80 $\pm$ 0.61 & 6.20 $\pm$ 0.80& 2.65 $\pm$ 0.90 & -0.74 $\pm$ 0.09& NAT \\
 
 & S3 & 03 12 56.9 & +08 22 11   & 3.25 $\pm$ 0.30 & 2.40 $\pm$ 0.60& --- & -0.40 $\pm$ 0.05 & unresolved \\

 & S4 & 03 12 57.5 & +08 22 09   & 137.17 $\pm$ 4.50 & 96.60 $\pm$ 5.90&33.50 $\pm$ 3.16 & -1.14 $\pm$ 0.05 & NAT \\
 
 & S5 & 03 12 57.3 & +08 21 37  & 8.78 $\pm$ 0.56 & 5.50 $\pm$ 0.70 & 2.82 $\pm$ 0.39 & -0.87 $\pm$ 0.09 & NAT \\

& S6 &  03 12 58.3& +08 21 14   & 65.72 $\pm$ 3.80 & 48.00 $\pm$ 4.20 & 19.63 $\pm$ 1.94 & -0.90$\pm$ 0.06 & NAT \\

& S7 & 03 12 54.4 & +08 20 35  & 2.50 $\pm$ 0.07 & 2.00 $\pm$ 0.09 & 1.50 $\pm$ 0.22 & -0.50 $\pm$ 0.07 & unresolved \\

 \hline

  \end{tabular}
  \label{ptsrc_tab}
  \tablecomments{The flux density values were estimated using the emission contained within the 3$\sigma$ region. The blank spaces are due to the difficulties of detecting those sources, either due to poor resolution or poor sensitivity.}
\end{table*}

\subsection{Continuum Images}

In Figure~\ref{img:1}, we have presented the uGMRT 400 MHz full-resolution continuum image of the PLCK171 cluster field. The central region is dominated by tailed galaxies (marked by the red crosses in Figure~\ref{img:1}) and the diffuse low surface brightness radio halo emission. The central radio halo emission is morphologically similar to the \textit{Chandra} X-ray thermal emission (see also in Figure~\ref{img:8a}). In the following sections, we will discuss the diffuse radio halo emission in the light of new images, highlighting new features.

\subsubsection{Radio halo emission}

\begin{figure*}
    \includegraphics[width=9.5cm, height = 9.5cm]{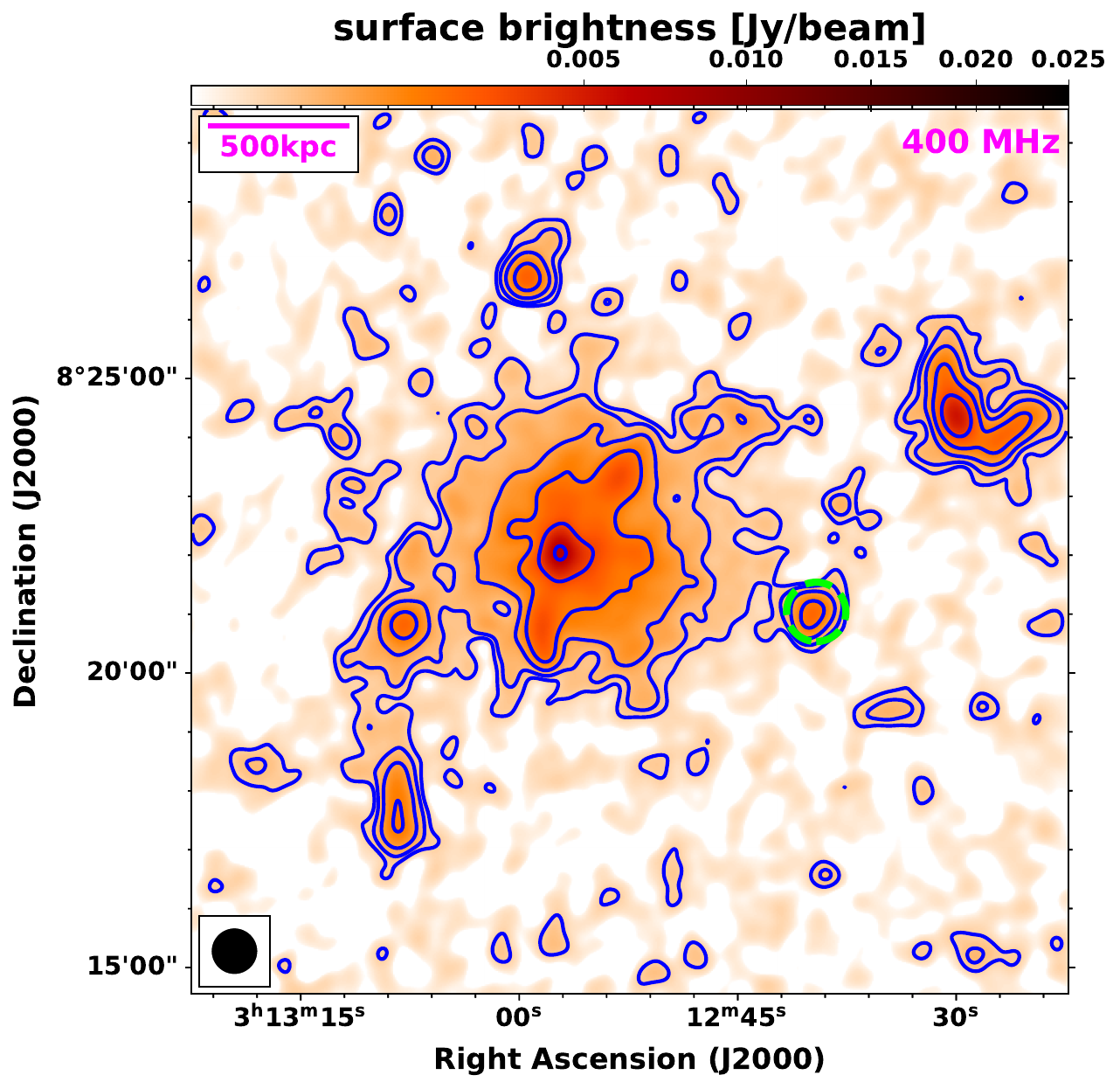}
    \includegraphics[width=8.55cm, height=9.5cm]{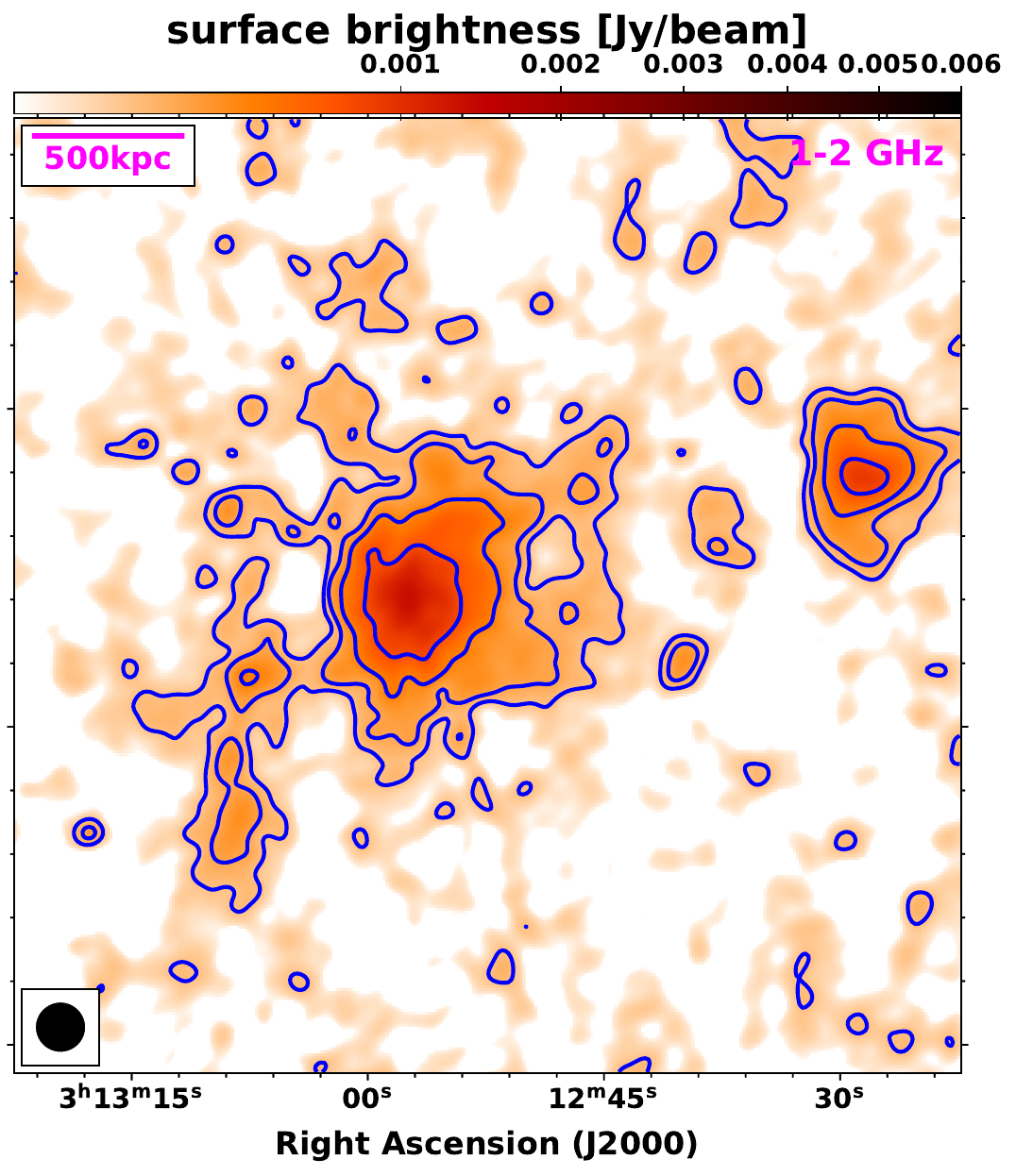}
    \caption{\textit{Left:} point source subtracted image of the cluster central field for 400 MHz at 25$''$ $\times$ 25$''$ (black circle at bottom left corner) is shown. The green circle indicates the detection of the source U at significance \textgreater 3$\sigma_{\rm rms}$. Contour levels are drawn at [1,2,4,8...]$\times$ 3$\sigma_{\rm rms}$, where $\sigma_{\rm rms}$ = 60 $\mu$Jy~beam$^{-1}$. \textit{Right:} Point source subtracted image at 1400 MHz, with a similar beam size. The RMS noise of the image is 30 $\mu$Jy~beam$^{-1}$ and the spacing of the contour level is similar to the left image.}
    \label{img:4a}
    \label{img:4b}
\end{figure*}

We have detected the extended and diffuse radio halo emission at frequencies 400 and 1400 MHz, shown in Figure~\ref{img:2a}. The overall morphology of the radio halo is circular, however, a slight elongation is observed along the southeast to the northwest axis, which is believed to be the merger projected direction \citep{2018ApJ...858...42A}. We have detected 7 discrete sources (\textgreater 4$\sigma_{\rm rms}$) in the halo region, out of them five are Narrow Angle Tailed (NAT) galaxies (shown in Figure~\ref{img:3}). The total flux density of these seven sources is $243.9 \pm 4.2$~mJy, which is $\sim$ 67 \% of the total flux density (radio halo $+$ individual sources) at 400 MHz (see Table~\ref{ptsrc_tab}). At 400 MHz, the radio halo covers an area of Largest Linear Size (LLS) of 1.50 $\times$ 1.30 Mpc$^{2}$ (Figure~\ref{img:4a}), whereas at 1400 MHz it is 1.20 $\times$ 1.03 Mpc$^{2}$ in the discrete source subtracted images. \citet{2013ApJ...766...18G} had reported the size of the radio halo to be $\sim$ 1 Mpc at 235 MHz. The radio halo is more extended at lower frequencies, and it is seen to be connected to the AGN in the southeast at 400 MHz (Figure~\ref{img:4a}, blue-lined sector). The radio halo brightness shows a significant decline in the outermost regions. The overall radio surface brightness is very smooth over the spatial scale of the radio halo at this resolution.   

\subsubsection{Radio galaxies in the region}

All the tailed galaxies have been found to have an optical counterpart in HST/ACS images (HST image is taken from \citealt{2018ApJ...858...42A}). We have overlaid the blue contours on the optical gray-scale image (Figure~\ref{img:3}) from 610 MHz high-resolution images to point out the optical counterparts of the tailed galaxies. Due to the small Field of view of the HST image, we have not marked the S7 in the image. The flux densities of these sources at different frequencies along with their positions are shown in Table ~\ref{ptsrc_tab}. The cluster center is very close (at a distance of 5$''$) to the source S4. Five sources (S1, S2, S4, S5, S6) out of these total 7 sources show an extended and bent-tail morphology, with in direction opposite to the cluster center. Given the limitation of the resolution of our available observations, it is very hard to comment on whether their jets are one-sided or two-sided, with a very narrow bending angle. Using the cluster redshifts, we have estimated the projected length of the NATs at 400 MHz in the uGMRT image, which ranges from $\sim$ 100 kpc (S2) to $\sim$ 220 kpc (S6), and their radio powers at 400 MHz of the order of $\sim$10$^{24}$--10$^{25}$ W~Hz$^{-1}$. The source S6 has a kink and another bend along the NE-SW directions up to $\sim$80 kpc, detecting only high-resolution uGMRT 400 MHz images, due to high sensitivity. These values are consistent with the range of size and radio power typically measured for NAT radio galaxies \citep[e.g.,][]{2017A&A...608A..58T}.

\subsubsection{About the source `U'}

We have detected a diffuse structure on the cluster periphery (Figure~\ref{img:4a}, green dashed circle), having a size of 75$''$ (310 kpc) and 53$''$ (219 kpc) respectively, at 400 MHz. It is situated at a projected distance of $\sim$ 900 kpc from the cluster center. This structure is also detected at a significant level at 1400 MHz, with a decrease in size with increasing frequency. We do not find any evidence of any optical counterpart for this source in the DSS2 (Digitized Sky Survey 2) image, and it is out of the HST Field of View. One of the possibilities arises of being a radio relic, due to its position at the cluster's periphery. However, the size of the source is much smaller than that of typical relics.

\begin{table}
  \centering
  \caption{Flux density estimates for diffuse sources}
\begin{tabular}{@{}cccc@{}}
    \hline
     Source & Freq. (MHz) & Flux density (mJy) & Ref.  \\
      \hline\hline
    
    Halo &235 & $253 \pm 27$ & J.21    \\

    &325 & $156 \pm 18$ & J.21   \\

    &402 & $115 \pm 14$ & This work \\
    
    &610 & $61\pm 9$ & J.21  \\

    &1400& $24.5\pm 4$ & This work  \\

    \hline

    Source U & 235 & $12.0 \pm 1.0$ & J.21      \\

   & 325 & $8.0 \pm 1.0$ & J.21 \\
   
    & 400 & $6.1 \pm 1.5$ & This work     \\
    
   & 1400 & $1.5 \pm 0.3$ & This work     \\

\hline
\end{tabular}
     \tablecomments{The flux density values were estimated using a common circular region from all the images. The reference J.21 refers to \citet{Joshi_21}.}
     
  \label{int_flux}
\end{table}

\subsection{Integrated spectrum} \label{int-spec}

Here, we combined our new uGMRT and JVLA observations with the legacy GMRT observations to investigate the spectral characteristics of the diffuse radio halo. We created the discrete source subtracted images with a common inner \textit{uv}-cut (as mentioned in \href{http://www.ncra.tifr.res.in/~ruta/files/Deepak-Joshi-MPhil-thesis.pdf} {J.21}) 
and convolved them into the same beam to ensure that we recover the flux density on the same spatial scales at all observed frequencies. The low-resolution point source subtracted image has been used to calculate the integrated flux density of the radio halo. We have reported the flux density values at different frequencies, in Table~\ref{int_flux}.

\begin{figure}
	\includegraphics[width=\columnwidth]{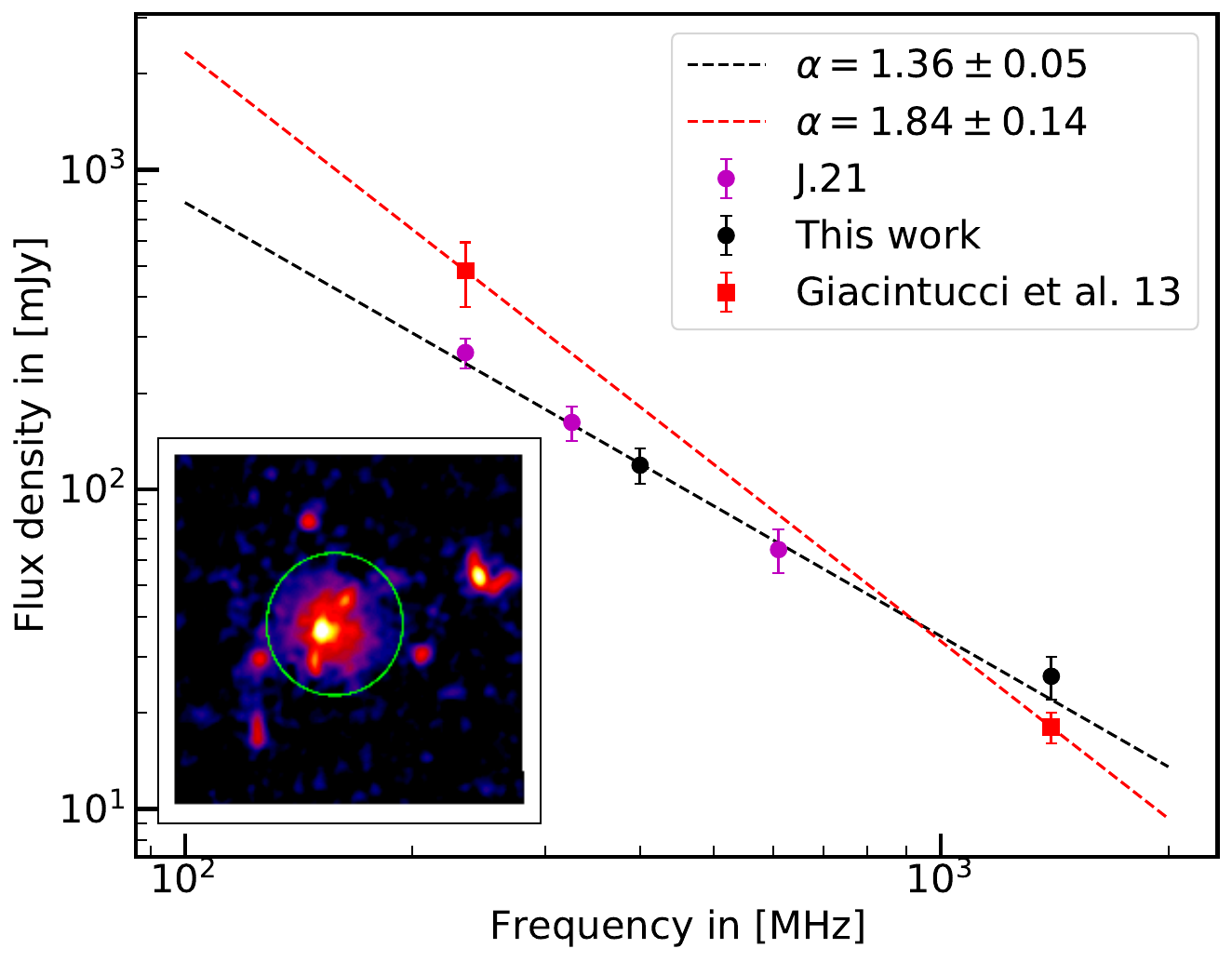}
         \caption{Integrated radio spectral index is shown for the radio halo between 235-1400 MHz. The black dashed line shows the overall spectrum of the halo following a single power law from our analysis. The black points denote the flux density values at each frequency in our work. The power law fit from the \citet{2013ApJ...766...18G} is shown for their 235 and 1400 Mhz measurements in red points. The annotated figure shows the 400 MHz radio halo emission (in color), and the green circle shows the radio halo region, which is used to estimate the flux density}
         \label{img:5}
\end{figure}

The overall spectrum of the radio halo is well fitted with the single power law between the frequency range of 235--1400 MHz, having an integrated radio spectral index of -1.36 $\pm$ 0.05 (Figure~\ref{img:5}). Therefore, a deeper on-source integration over a wide frequency range indicates that the radio halo in the PLCK171 is not an ultra-steep spectrum radio halo. \citet{Joshi_21} had also shown the spectral index of the halo to be -1.27 $\pm$ 0.08, using 235, 325, and 1400 MHz (VLA; D-array) observation, matches well with our analysis within 1$\sigma$. We have also investigated the 3 GHz VLASS images to obtain the radio halo flux density, however, those data sets were heavily affected by the calibration errors, also the VLA B-array does not properly sample large angular scale emission. Using our multi-frequency observations, spanning over 235$-$1400 MHz, we have also estimated the integrated spectrum of source U, coming out to be -1.15 $\pm$ 0.02.

The comparison between our work and \cite{2013ApJ...766...18G} demonstrated the significant change in the flux density at 235 and 1400 MHz. The flux density values of the tailed galaxies at 235 MHz are higher (e.g., the combined flux density of the S3 and S4 is 292.3 mJy in our work, and 248.4 mJy in \citealt{2013ApJ...766...18G}, and significant difference in the flux density of the source S6). Therefore, the subtracted flux density in the new 235 MHz image might be higher, as the individual galaxies have higher flux. The number of sources (S1$-$S6) subtracted in \citet{2013ApJ...766...18G}, and the new image is similar, however, the discrepancy lies in the amount of subtraction of the flux densities. The NVSS pointing was very short ($\sim$ 15 mins) on source time, making it very hard to recover the diffuse emission at shorter baselines. Our C$+$D array analysis has been able to map the extended emission up to larger radii as well as better subtraction of the contribution from the discrete sources, implying higher flux for the extended radio halo emission at 1.4 GHz.

We have used the flux densities measured at 1.4 GHz for the estimation of the total radio power of the halo. The total rest-frame radio power of the halo is P$_{1.4 \rm GHz}$ $\sim$ 8 $\times$ 10$^{24}$ W Hz$^{-1}$. This radio power sits well (within the 7.6\% of the expected value) in the P$_{1.4 \rm GHz}$--Mass and P$_{1.4 \rm GHz}$--L$_{\rm X}$ planes for giant radio halos \citep{2013ApJ...777..141C, 2021A&A...647A..51C}.

\subsection{Radio surface brightness profile}

\begin{figure}
	\includegraphics[width=\columnwidth]{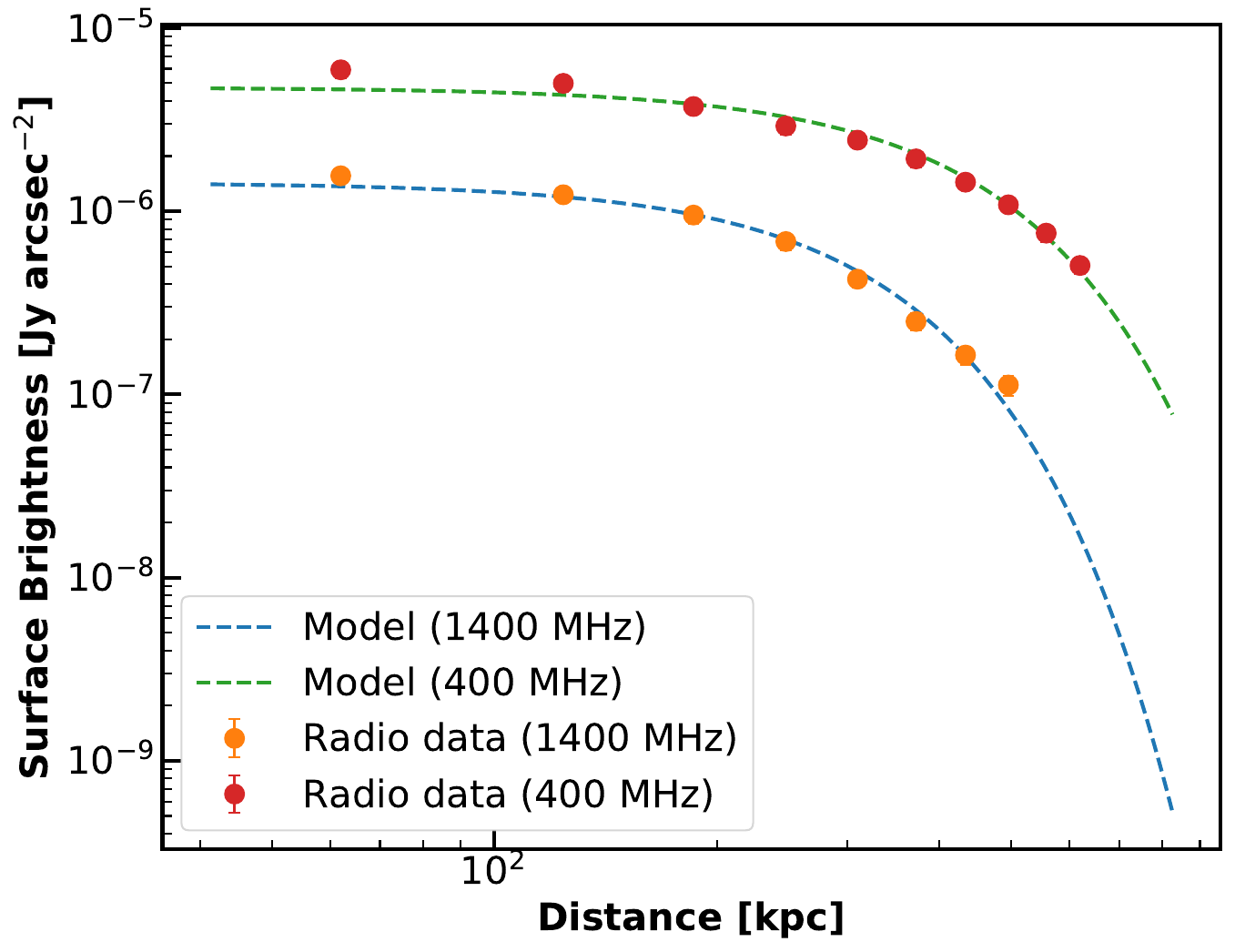}
         \caption{The radial profile of the radio surface brightness is shown. Each of the circles (different colors at different frequencies) denotes the azimuthally averaged radio surface brightness values, within the annular regions. The error bars represent the error on the mean. The dashed line shows the two-dimensional model fitted to the radio surface brightness profile.}
         \label{img:6}
\end{figure}

We have also fitted the surface brightness profile of the halo and checked how the averaged radio surface brightness varies with distance. We have adopted the approach by \cite{2009A&A...499..679M} and very recently by \cite{2021A&C....3500464B}, where one can model the surface brightness variation by circular model, as well as taking into account the asymmetric and elongated shape of the radio halo. The surface brightness profile is given by:

\begin{equation}
    \rm I(r) = \rm I_{0} \exp^{-\frac{r}{r_{e}}}
\end{equation}

where I$_{0}$ is the central surface brightness, and r$_{e}$ denotes the e-folding radius, the distance at which the surface brightness falls to a value of I$_{0}$/e. In Figure~\ref{img:6}, we have shown the best fit to data. The primary advantage of the \texttt{HALO-FDCA} \citep{2021A&C....3500464B} procedure is that the profiles are fitted to a two-dimensional image directly, using the \texttt{MCMC} (Markov Chain Monte Carlo) chain to estimate the parameters. The profile has been obtained by averaging the radio brightness in concentric annuli. The central annular region is drawn at 25$''$, and the cluster center is chosen to be the center of the annuli. The central radio brightness is $\sim$ 8.5 $\mu$Jy~arcsec$^{-2}$, with an e-folding radius of 280 kpc. In Table ~\ref{radio_srfce_brtnss}, we have reported our results along with the values obtained at different frequencies. We note that the e-folding radius at JVLA frequency is slightly smaller than at uGMRT frequency, indicating a more peaked profile of the radio emission. However, the e-folding radius is very close to each other within 1 $\sigma$. \citet{2021A&A...647A..51C} had also reported the e-folding radius and central brightness at 1.4 GHz using the NVSS D-array images, and their e-folding radius matches with ours within the error bars.

\begin{table}
  \centering
  \caption{\texttt{HALO-FDCA} fitting parameters}
  \begin{tabular}{@{}ccccc@{}}
    \hline
      Freq. (MHz) & $\chi_{r}^{2}$ & I$_{0}$ & r$_{\rm e}$ & Ref.  \\
    \hline\hline

    400 & 1.76&8.54 $\pm$ 0.04 & 280 $\pm$ 9 & This work \\

    1400 & 2.15 & 2.11$\pm$ 0.12 & 263$\pm$ 11  & This work   \\

    1400 (D array) & 2.23 &1.35$\pm$ 0.11 & 253$\pm$ 15  & C.21   \\

    \hline
  \end{tabular}
  \label{radio_srfce_brtnss}
  \tablecomments{Column (1): Frequency used for the analysis. Column (2): reduced $\chi^{2}$ value. Column (3): central brightness of the fit in units of $\mu$Jyarcsec$^{-2}$. Columns (4): e-folding radii units of kpc. The reference C.21: \citet{2021A&A...647A..51C}}
\end{table}

\begin{figure*}[t!]
    \includegraphics[width=9.5cm, height = 8.3cm]{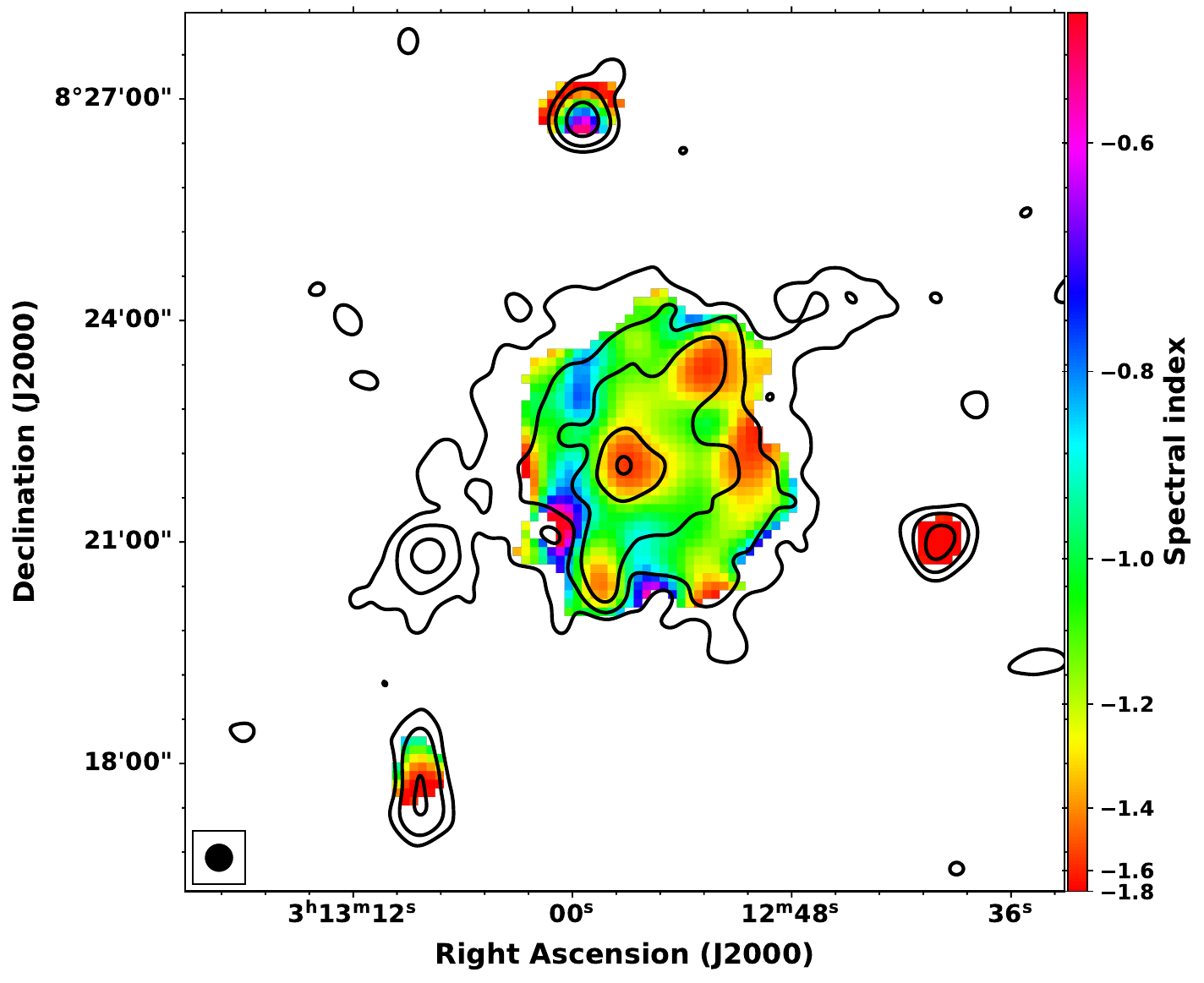}
    \includegraphics[width=8.6cm, height=8.3cm]{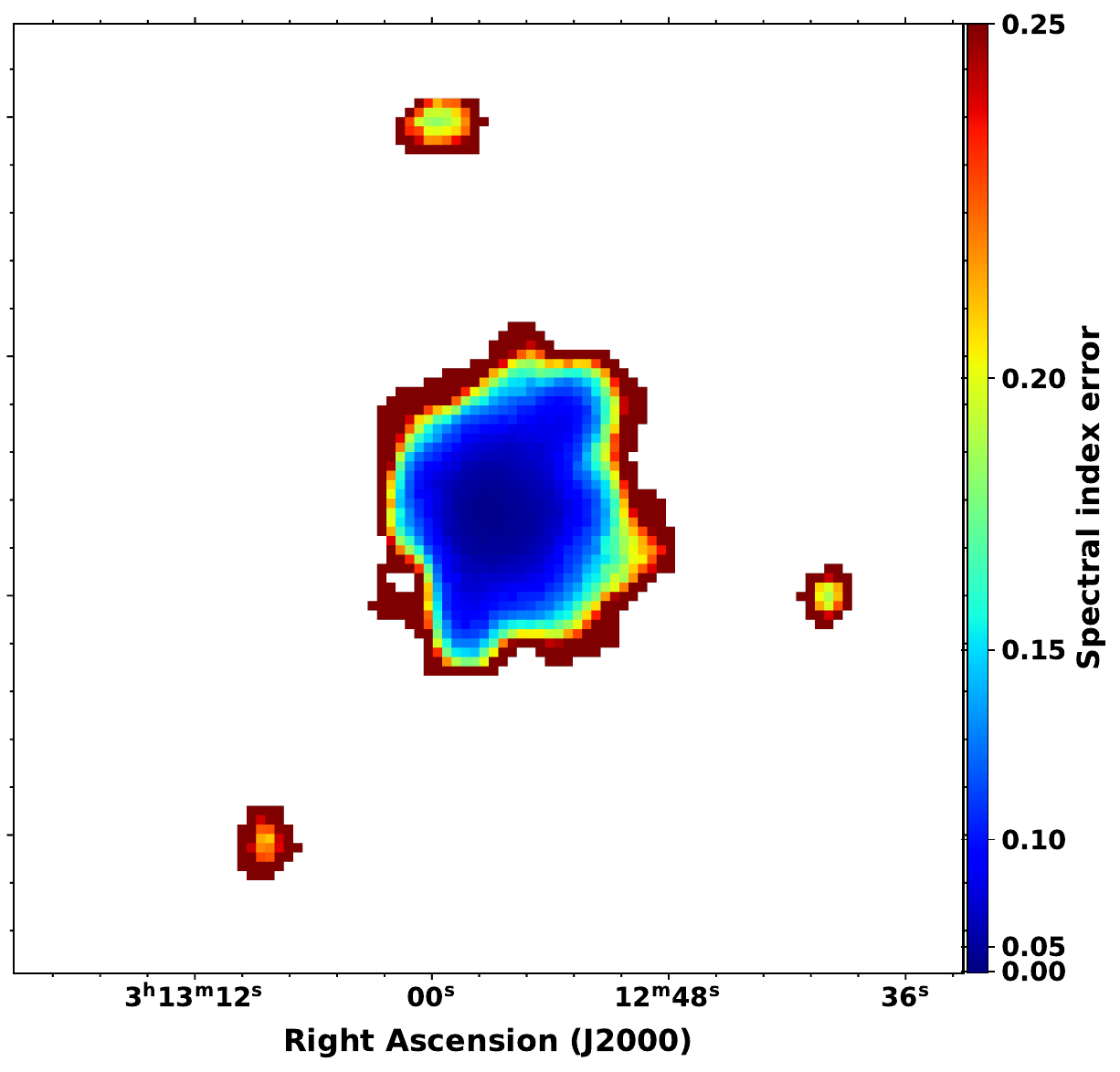}
     
    \caption{\textit{Left:} Spectral index map for the radio halo in PLCK171 at 25$''$ is shown between 400 and 1400 MHz. The black contours are drawn at 3$\sigma_{\rm rms}$ $\times$ [1, 2, 4, 8..] from the 400 MHz image. \textit{Right:} The corresponding error map is shown here. The procedures for the estimation of the error map are followed as described in \cite{2017SciA....3E1634D}.}
    \label{img:7a}
    \label{img:7b}
    
\end{figure*}

\subsection{Spectral index map}

Spatially resolved spectral maps are a tool to understand the complex physics at local ICM scales \citep{2016ApJ...818..204V}. Our uGMRT 400 MHz and JVLA 1400 GHz images have allowed us to study the resolved spectral map for the radio halo. Despite the moderate resolution of the point source subtracted image, the resolved spectral index map for the diffuse emission is obtained with a satisfactory S/N.

The spectral index maps were created using the point source subtracted visibilities at 400 and 1400 MHz, with a similar inner \textit{uv} cut-off and uniform weighting to match the spatial scale of the radio emission. Then both images were convolved to a common resolution of 25$''$, such that the radio halo emission could be highlighted properly at both frequencies. The resolution of the image is chosen such that there is an optimization between probing the spectral variation in the radio-emitting region and having a good signal-to-noise ratio. We have followed the procedure, described in \cite{2017SciA....3E1634D}, for making the spectral index map. For each image, the pixel value of less than 3$\sigma_{\rm rms}$ is masked out.

Figure~\ref{img:7a} shows the resolved spectral index image (left panel) and corresponding error map (right panel) for the radio halo. In general, the spectral index is quite uniform around the values ranging from -1.0 to -1.5. However, some steeper regions are visible, although most of them can be associated with some residual emission from Head-Tail radio sources, known to display very steep spectra in their outer parts. The source 'U' shows a very steep spectral index of $\sim$ -1.60, with an error of 0.25, very different from the integrated spectrum obtained in sec.~\ref{int-spec}. Since the detection of the `U' in the VLA image is not very extended, therefore, it is very hard to conclude whether the spectral index is steep throughout the extent of the source. The errors corresponding to the spectral indices at the central regions are below 0.1. Any visible fluctuations in the spectral map throughout the radio halo may occur due to the patchy distribution of the magnetic field and various acceleration efficiencies at different scales \citep[e.g.,][]{2001MNRAS.320..365B,2007MNRAS.378..245B}. However, artificial patches in a resolved spectral index map can also originate due to unmatched \textit{uv}-coverage in the different frequencies. We have also checked the spectral index as a function of distance from the cluster center by sampling the total radio halo into beam-sized annular regions. We do not find any particular trend (steepening or flattening) with distance from the center. We emphasize that such radial invariance of the spectral index may originate due to the intrinsic physics of the ICM; however, at this low resolution, the fluctuations may have been averaged out.

\begin{figure}
	\includegraphics[width=\columnwidth]{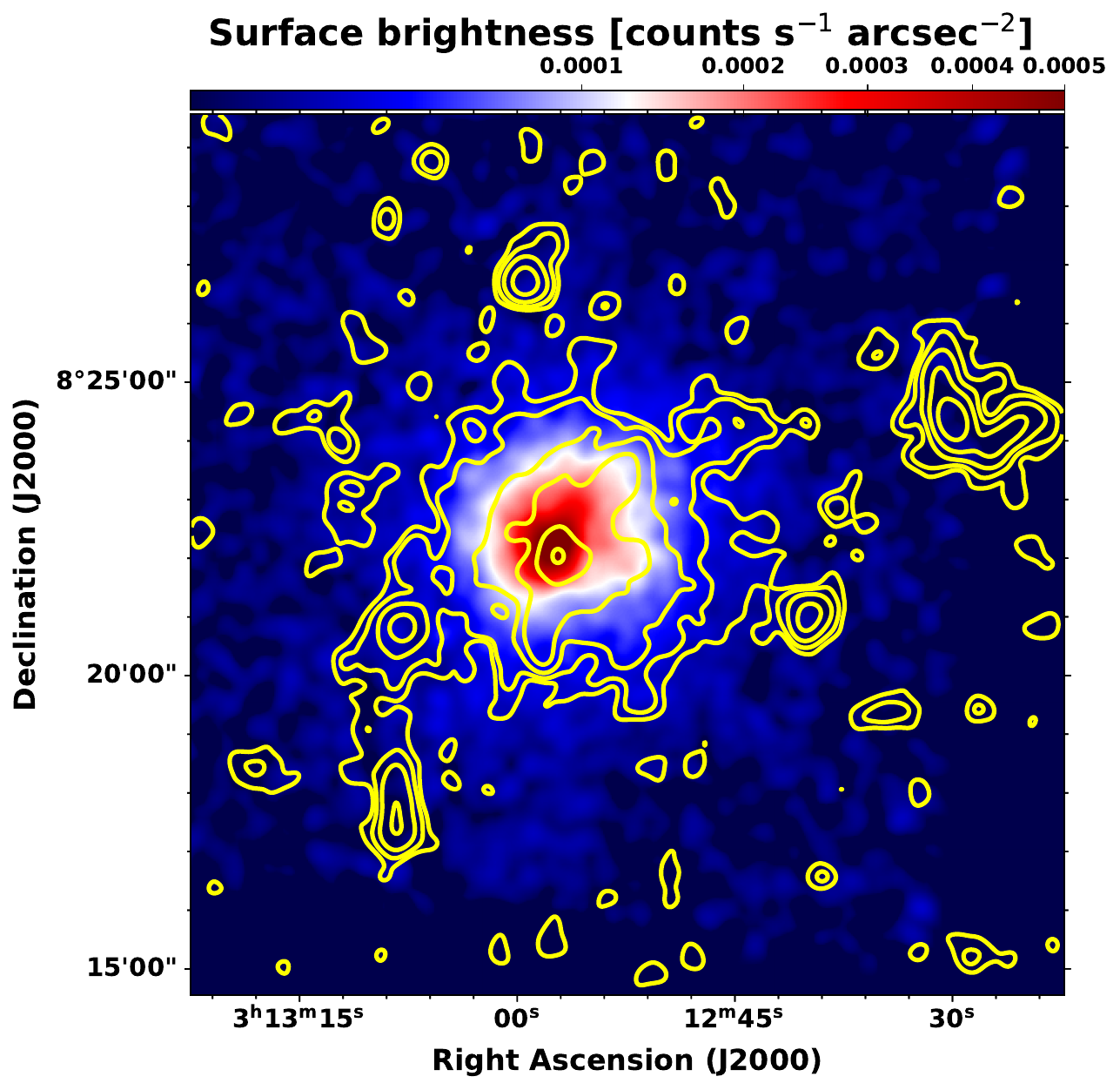}
         \caption{A \textit{Chandra} image (in color) of PLCK171 in the 0.5$-$4.0 keV band is shown, smoothed with a Gaussian Full-width half maximum of 5$''$. The X-ray image is background subtracted, divided by the exposure map, and the point source removed. The radio emission at 400 MHz is overlaid as yellow contours. The contours level starts from 3$\sigma_{\rm rms}$ $\times$ [1, 2, 4..], where $\sigma_{\rm rms}$ = 60 $\mu$Jy~beam$^{-1}$. }
         \label{img:8a}
\end{figure}

\section{Interplay of thermal gas and non-thermal plasma} \label{sec:4}
 
Theoretical models for the formation of radio halos would expect that radio emission approximately follows the X-ray emission from the thermal gas and this expectation has been established via deep and sensitive observations over the years \citep{2001A&A...369..441G, 2011MNRAS.412....2B}. However, there are a few clusters whose radio emission does not trace the thermal emission with the same morphology \citep{2005A&A...440..867G, 2018MNRAS.473.3536W, 2018MNRAS.478.2927B}. This Morphological resemblance between the X-ray and radio can be used as a fundamental probe to study interactions and different physical processes between the thermal and non-thermal plasma of the ICM.

We present the \textit{Chandra} X-ray surface brightness map of the PLCK171, overlaid with radio emission (yellow contours) at 400 MHz (Figure ~\ref{img:8a}). The radio emission from the halo is co-spatial with the diffuse thermal X-ray emission. The thermal emission fades at the outer distance. We do not see any significant X-ray emission associated with the source `U'. The overall geometry of the thermal emission is circular, however, in the central regions, the surface brightness seems to be slightly elongated along the merger axis (northwest-southeast), suggested by the \textit{XMM-Newton} gas temperature map derived by \citet{2013ApJ...766...18G}.

\subsection{Point-to-point analysis}

In the scenario of turbulent acceleration, a small portion of the turbulent energy is transferred to the non-thermal plasma of the ICM, accelerating the charged particles to relativistic velocities and amplifying the seed magnetic field. The rest of the energy is converted into thermal energy within the ICM, increasing the average temperature of the ICM. Therefore, a correlation study between the radio and X-ray surface brightness can constrain the radio halo origin models \citep{2001A&A...369..441G}. To understand the correlation between the radio and X-ray brightness, we have used discrete source subtracted radio images created at 25$''$ resolution. The \textit{Chandra} X-ray image was smoothed with a Gaussian of 5$''$ full width at half maximum (FWHM). We have followed the procedure mentioned in \cite{2020A&A...640A..37I, 2022NewA...9201732I} to sample the radio and X-ray emissions. With the high sensitivity of the available uGMRT, JVLA, and \textit{Chandra} X-ray data, we have performed a detailed investigation of the point-to-point radio and X-ray correlation at 400 MHz, and 1.4 GHz. The radio brightness (I$_{\rm R}$) is quoted in the units of Jy~arcsec$^{-2}$ and X-ray (I$_{\rm X}$) in units of Counts~s$^{-1}$~arcsec$^{-2}$.

\begin{figure}
	\includegraphics[width=\columnwidth]{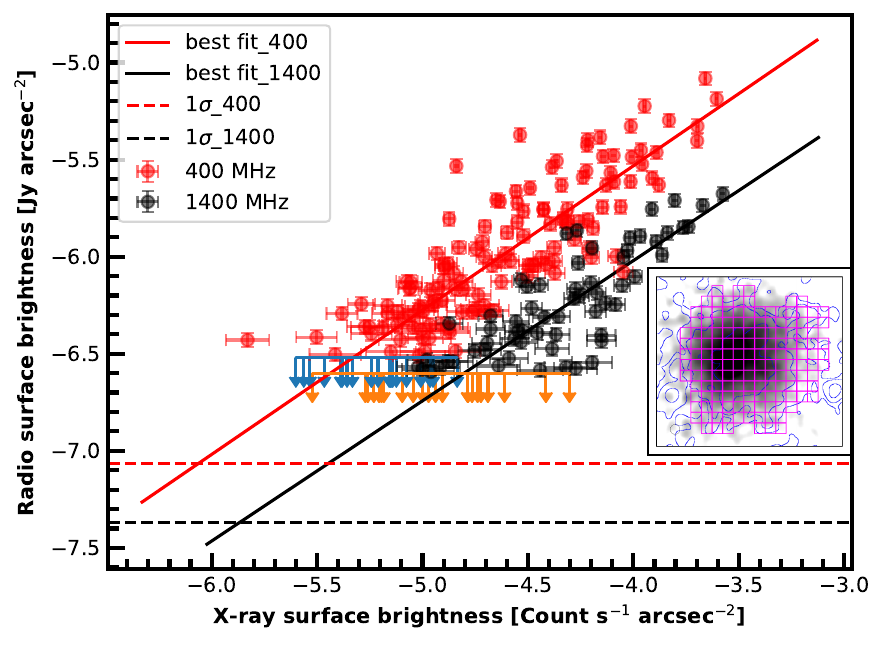}
        \caption{Radio vs X-ray surface brightness (on a logarithmic scale) plots for the halo in PLCK171 at 400 and 1400 MHz are shown. The LinMix best-fit line is shown in the solid (red \& black) lines. The red and black circles show the detection of the radio halo above 3$\sigma_{\rm rms}$. The 2$\sigma_{\rm rms}$ are considered as the upper limits and indicated by arrows (cyan for 400 and orange for 1400 MHz). The red and black dashed line shows the 1$\sigma_{\rm rms}$ line for different frequencies. The inset image shows the diffuse X-ray emission in gray color, and radio emission in blue contours. The magenta grids are (25$''$) the regions used for estimating the radio and X-ray surface brightness.} 
    \label{img:9}
\end{figure}

Figure ~\ref{img:9} shows the comparison between the X-ray and radio brightness in the log-log scale for both frequencies over the extent of the halo. The thermal and non-thermal emission components of the halo show a clear positive correlation, the high radio brightness corresponding to high X-ray brightness regions. Several previous radio halo studies have reported a relationship between the radio and X-ray brightness \citep[e.g.,][]{2014MNRAS.440.2901S, 2019A&A...628A..83C, 2020ApJ...897...93B, 2023A&A...669A...1R,2023A&A...678A.133B, 2024ApJ...962...40S,2024arXiv240300414R,2024arXiv240218654B}. This relationship is generally described by a power law of the form:

\begin{equation}
    \log(\rm I_{\rm R}) = a + b\log(\rm I_{\rm X})
\end{equation}

where a slope of b = 1 suggests a linear relation, indicating that the radial decay of the relativistic electrons and magnetic field of the ICM, is similar to the thermal gas, and b \textless 1 (sublinear) implies a slower spatial change of non-thermal components compared to thermal components or vice versa (if b \textgreater 1).  

\begin{table}
  \centering
  \caption{Linmix fitting slopes correlation coefficients of the data for Figure~\ref{img:9}.}
  \begin{tabular}{@{}ccccc@{}}
    \hline
      $\nu$ & b$_{\rm upperlimits}$ & b & r$_{\rm s}$ & r$_{\rm p}$  \\
    \hline\hline

    400 & 0.72 $\pm$ 0.03 & 0.68$\pm$ 0.05 & 0.90 & 0.89 \\

    1400 & 0.71 $\pm$ 0.04 & 0.64$\pm$ 0.04 & 0.88 & 0.83  \\

    \hline
  \end{tabular}
  \label{corr_res}
  \tablecomments{Column (1): Frequency used for the analysis. Column (2): correlation slope including the upper limits (3): correlation slope without upper limits (4): Spearman coefficient (5): Pearson coefficient}
\end{table}

To quantify the strength of the correlation, and to obtain the correlation slope to the observed data of the I$_{\rm R}$ - I$_{\rm X}$ relations, we adopt the Linmix\footnote{For more information on LinMix check \url{https://linmix.readthedocs.io/en/latest/src/linmix.html}} package \citep{2007ApJ...665.1489K}. Linmix does a Bayesian linear regression and accounts for the measurement uncertainties on both variables, intrinsic scatter, and upper limits. The correlation strength was measured by using the Spearman (r$_{\rm s}$ and Pearson correlation (r$_{\rm p}$) coefficients. We summarize the best-fit slopes and corresponding correlation coefficients for each radio frequency in Table~\ref{corr_res}. We find that the radio and X-ray surface brightness have a strong correlation at both frequencies. At 400 MHz, the slope b$_{400 \rm MHz}$ = 0.72 $\pm$ 0.03, and shows low intrinsic scatter of $\sigma_{\rm int}$ = 0.181 $\pm$ 0.006, whereas at 1400 MHz, the slope is b$_{1400 \rm MHz}$ = 0.71$\pm$ 0.04, with an intrinsic scatter of 0.162$\pm$0.004. The above values are in line with the slopes obtained for the giant radio halos.

The frequency dependence of the correlation slope has been studied for many radio halos, particularly in the cases of Abell 520 \citep{2019A&A...622A..20H} and MACSJ017.5+3745 \citep{2021A&A...646A.135R}, where the investigations for the I$_{\rm R}$ $-$ I$_{\rm X}$ correlations were done for multiple frequencies. They report a significant variation of correlation slope over frequency, indicating spectral steepening at high frequencies. However, there are cases where the slope remains constant as a function of frequency: Abell 2744 \citep{2021A&A...654A..41R}, Abell 2256 \citep{2023A&A...669A...1R}. In PLCK171 we also found that the correlation slope is nearly constant across such a wide frequency range of 1 GHz.

\section{Discussions} \label{sec:5}

\subsection{The tailed radio galaxy}

We have used the \textit{Chandra} observation to study the environment by analyzing the ICM at the location of the head-tail galaxies. Disturbed clusters tend to bend the jets of the radio galaxies via ram-pressure and other physical processes: each host galaxy moves with respect to the ICM \citep[e.g][]{2017A&A...608A..58T, 2018A&A...609A..61C, 2020MNRAS.493.3811S, 2023A&A...675A.118I}. To understand the ambient ICM environment of the radio galaxies under study, we used the X-ray information about the cluster gas distribution by assuming that it can be approximated by a spherically symmetric double $\beta$-model \citep{1976A&A....49..137C}:
\begin{equation}
    n(r) = \frac{n_{0}}{1+f}\left[\left(1 + \frac{\rm r^{2}}{\rm r_{\rm c1}^{2}}\right)^{-\frac{3}{2}\beta_{1}} + \left(1 + \frac{\rm r^{2}}{\rm r_{\rm c2}^{2}}\right)^{-\frac{3}{2}\beta_{2}}\right]\, , 
\end{equation}
where \textit{n(r)} is the electron density at a distance r from the cluster center, \textit{n$_{0}$} is the gas density at the center, and r$_{\rm c1}$, r$_{\rm c2}$, $\beta_{1}$, $\beta_{2}$, $f$ are the free parameters. Since the true distance to the host galaxies from the cluster center is unknown, we use the projected distance, and hence, the estimated densities should be considered as upper limits. Considering the projected distance of the radio galaxies from the cluster center, the $\beta$ model yields electron densities from a range of (3.3--9.0) $\times$ 10$^{-3}$ cm$^{-3}$. (The model parameters are taken from \citealt{2017ApJ...841...71G} for this cluster.) The results indicate that the density of the ICM at the radio galaxy location can be very high, justifying the hypothesis that the bent structure may occur due to the interaction of high-velocity host galaxies moving through the dense ICM. We also want to emphasize that the above calculation is very simplistic and based upon a lot of assumptions. Using the information of temperatures from \citet{2013ApJ...766...18G}, the gas pressure via P$_{\rm gas}$ = 1.83 n$_{e}$k$_{\rm B}$T, the gas density $\rho$ = 1.9~$\mu$n$_{e}$m$_{H}$ with the mean molecular weight $\mu$ $\approx$ 0.61 and the proton rest mass m$_{H}$, and the speed of sound = $\sqrt{\frac{\gamma P}{\rho}}$, where $\gamma$ = 5/3 is the adiabatic index for a mono-atomic, thermal gas can be obtained. The obtained pressure at the location of the head-tail galaxies ranges from (5.38--12.60) $\times$ 10$^{-11}$ erg cm$^{-3}$, and the gas density ranges from (9.8--18.0) $\times$ 10$^{-27}$ g cm$^{-3}$. We assume that the galaxies are moving velocities of order $\sim$ 1500 km~s$^{-1}$ with respect to the cluster center (corresponding to the cluster three-dimensional velocity dispersion, which is estimated using the virial theorem), and this assumption is used to calculate the sonic Mach numbers M$_{s}$. The obtained sound speed ranges from 1214--1410 km~s$^{-1}$. All galaxies are found to move supersonically through the ICM (M$_{s}$ \textgreater 1.1--1.0). The galaxies experience a ram pressure of (1.3--3.0) $\times$ 10$^{-10}$ g cm$^{-1}$~s$^{-2}$, causing the bending of the galaxy jets in the outward direction of the cluster center.

\subsection{Dynamics of the merger}

\citet{2017ApJ...846...51L} have carried out a detailed X-ray morphological analysis of a large sample (189) of Planck ESZ clusters, including PLCK171 using XMM-Newton images. PLCK171 was reported to be a relaxed cluster from their analysis, with a concentration of 0.17 $\pm$ 0.01, and a centroid shift of (1.14 $\pm$ 0.07) $\times$ 10$^{-2}$. However, different values have been obtained for the centroid shift (2.3 $\times$ 10$^{-2}$), and concentration parameter (0.14) from the \textit{Chandra} data by \cite{2021A&A...647A..51C}, and they have reported it as a merging cluster. The observed asymmetry of the ICM at the central region along the NW-SE directions suggested that a merger occurred along this axis. \cite{2017ApJ...841...71G} had estimated the value of the average temperature and central specific entropy floor of this cluster to be 11.6 keV, 329 $\pm$ 74 respectively, stipulating it as a consequence of high disturbance at the center. The evidence of the merging activity is also prominent due to the elongated mass distribution in the strong Lensing map using HST data \citep{2018ApJ...858...42A}. The X-ray brightness peak and the BCG's positions do not have significant offset along the line of sight, indicating a post-merger scenario, where the structure has retained its gas content after the pericentric passage of the subcluster. 

Several radio halo observations support the turbulent re-acceleration model, including the high-frequency spectral steepening, spectral index fluctuations across its spatial extent, and a strong sub-linear correlation between radio and X-ray surface brightness \citep[e.g.,][]{2020A&A...636A...3X,2021A&A...646A.135R,2021A&A...650A..44B}. Our multi-frequency radio analysis does not provide any indications of the high-frequency steepening, although our observations only span up to 1.4 GHz. The resolved spectral map overall indicates a uniform distribution, along with some patches of steep spectral region. The strong sub-linear positive correlation between I$_{\rm R}$ and I$_{\rm X}$ falls in line with the predictions from turbulent re-acceleration, as, in the hadronic origin scenario of the giant radio halos, a significantly steeper slope is expected depending on the magnetic field value \citep[e.g.,][]{2001A&A...369..441G, pfrommer2008simulating}. The estimated e-folding radius from the radio surface brightness profile shows a slight change over frequency, however, the values match within 1$\sigma$. A change in the e-folding radius signposts the radio brightness at two frequencies declines with different rates, hence a change in the correlation slope over frequencies, and radial spectral steepening. \citet{2022ApJ...933..218B} had reported a significantly different e-folding radius at 144 and 1314 MHz for the coma cluster. Our current radio results indicate towards the turbulent re-acceleration model as the primary origin for the radio halo, where the relativistic charged particles (possibly originated from the lobes of the NATs; \citealt{2024Galax..12...19V}) may have been re-energized via turbulence injected due to the violent merger. 
\citet{2017ApJ...843L..29E} has reported a correlation between the radio power (at 1.4 GHz) and the turbulent velocity dispersion. PLCK171 has been reported to have a high turbulent velocity dispersion of 438 $\pm$ 40 km~s$^{-1}$, and 1D Mach number of 0.15 $\pm$ 0.01, indicating a significant presence of the turbulence in the ICM.   

\section{Summary} \label{sec:6}

Here, we present the first detailed analysis of the radio halo in the galaxy cluster PLCK171 using multi-frequency observations. Earlier works presented this cluster with an ultra-steep spectrum radio halo using data only from legacy GMRT (235 MHz) and NVSS VLA (1.4 GHz). Here we have presented deeper observations from 235 MHz to 1400 MHz to verify the steep spectrum candidacy and understand the underlying particle acceleration at the cluster region. Our radio observations have been combined with available X-ray observations (\textit{Chandra}), providing strong indications of the thermal and non-thermal connections in the ICM. Our overall findings are summarised below:
 
\begin{enumerate}
     \item We present the high to moderate resolution (5$''-25''$) uGMRT and JVLA images of the galaxy cluster PLCK171. The radio halo emission in the central region has been recovered, with the extent of the radio halo being 1.5 Mpc at 400 MHz and 1.2 Mpc at 1400 MHz, respectively. Our new observations have discovered an unknown diffuse source `U' at $\sim$ 900 kpc away to the right side of the cluster center, with an LLS of 300 kpc.

\item  The radio halo integrated spectrum follows a single power law spanning a frequency range of 253 to 1400 MHz with a spectral index of -1.36 $\pm$ 0.05, establishing that it is not an ultra-steep spectrum radio halo. The halo fits well with the known radio power at 1.4 GHz $-$ cluster mass (M$_{500}$) correlation of other halos at 1.4 GHz. We have also obtained the integrated spectral index of the source `U' to be -1.15 $\pm$ 0.02, indicating a possibility of it being a radio relic. However, the current depth of the X-ray and radio observations prevents further conclusions from being drawn about that source.

\item The spatially-resolved spectral index (at resolution 25$''$) suggests overall a smooth distribution ($\sim$ -1.3) across the extent, with some patches of steep spectral index (-1.60 $\pm$ 0.05 ) regions. The error map shows that in the central region, the error is below 0.10. We do not find any specific trend of spectral index along the merging axis. We have also checked the radial variation of the spectral index, however, no steepening has been seen with the increasing distance along outwards. 

\item The non-thermal emission from the radio halo is found to be co-spatial with the ICM thermal emission, indicating that the hot gas and non-thermal plasma have a close correlation. A significant elongation along the merger axis has been observed in both X-ray and radio surface brightness maps. The occurrence of the merger is also strengthened by the presence of multiple-tailed galaxies, which are experiencing ram-pressure stripping at such dense cluster environments.

\item A strong positive sub-linear correlation, with a slope of $\sim$ 0.71 between the radio and X-ray surface brightness has been reported via a point-to-point analysis across the spatial extent of the radio halo, indicating turbulent re-acceleration as the driving mechanism for the origin of the halo. The slope was found to be relatively constant across the frequencies used for our study. 

\end{enumerate}

The highly disturbed galaxy cluster PLCK171 is a dynamically rich system that hosts a giant radio halo and many extended tailed galaxies. These extended radio sources are the classic test-bed for studying turbulence, particle (re-)acceleration, and magnetic field evolution in the post-merger stage of the cluster. Our current multi-frequency study provides new insights into the nature of the sources (i.e., morphology, spectral properties, and the interplay between thermal and non-thermal components). However, due to the quality of the current data, we are unable to confirm some aspects associated with, e.g., the classification of the unknown diffuse source, and any significant fluctuations in the spectral map to study the in-homogeneity of the emitting local ICM regions. Further work using optical weak-lensing observations, additional X-ray observations, and deep radio observations below 200 MHz will be necessary to investigate the merging scenario further.


\begin{acknowledgments}
We thank the anonymous referee for their constructive comments
that have improved the clarity of the paper. R.S. acknowledges Sameer Tanaji Salunkhe for many useful discussions. R.S. also thanks Gabriella Di Gennaro and Luca Bruno for their prompt replies to many useful queries. R.S. and R.K. acknowledge the support of the Department of Atomic Energy, Government of India, under project no. 12-R\&D-TFR-5.02-0700. R.K. also acknowledges the support from the SERB Women Excellence Award WEA/2021/000008. S.G. acknowledges that the basic research in radio astronomy at the Naval Research Laboratory is supported by 6.1 Base funding. We thank the staff of the GMRT that made these observations possible. The GMRT is run by the National Centre for Radio Astrophysics (NCRA) of the Tata Institute of Fundamental Research (TIFR). The scientific results reported in this article are based in part on data obtained from the \textit{Chandra} Data Archive (Obs ID 15302). This paper employs a Chandra dataset, obtained by the Chandra X-ray Observatory, contained in~\dataset[DOI:290]{https://doi.org/10.25574/cdc.290}. The National Radio Astronomy Observatory is a facility of the National Science Foundation operated under a cooperative agreement by Associated Universities, Inc. This research made use of the NASA/IPAC Extragalactic Database (NED), which is operated by the Jet Propulsion Laboratory, California Institute of Technology, under contract with the National Aeronautics and Space Administration. 
\end{acknowledgments}

%

\vspace{5mm}
\facilities{upgraded Giant Metrewave Radio Telescope (uGMRT), Karl G. Jansky Very Large Array (JVLA), \textit{Chandra}}


\software{astropy \citep{2013A&A...558A..33A,2018AJ....156..123A}, CASA, numpy, aplpy, matplotlib, ds9, CIAO}







\bibliography{sample631}{}

\begin{thebibliography}{}
\expandafter\ifx\csname natexlab\endcsname\relax\def\natexlab#1{#1}\fi
\providecommand{\url}[1]{\href{#1}{#1}}
\providecommand{\dodoi}[1]{doi:~\href{http://doi.org/#1}{\nolinkurl{#1}}}
\providecommand{\doeprint}[1]{\href{http://ascl.net/#1}{\nolinkurl{http://ascl.net/#1}}}
\providecommand{\doarXiv}[1]{\href{https://arxiv.org/abs/#1}{\nolinkurl{https://arxiv.org/abs/#1}}}

\bibitem[{{Acebron} {et~al.}(2018){Acebron}, {Cibirka}, {Zitrin}, {Coe}, {Agulli}, {Sharon}, {Brada{\v{c}}}, {Frye}, {Livermore}, {Mahler}, {Salmon}, {Umetsu}, {Bradley}, {Andrade-Santos}, {Avila}, {Carrasco}, {Cerny}, {Czakon}, {Dawson}, {Hoag}, {Huang}, {Johnson}, {Jones}, {Kikuchihara}, {Lam}, {Lovisari}, {Mainali}, {Oesch}, {Ogaz}, {Ouchi}, {Past}, {Paterno-Mahler}, {Peterson}, {Ryan}, {Sendra-Server}, {Stark}, {Strait}, {Toft}, {Trenti}, \& {Vulcani}}]{2018ApJ...858...42A}
{Acebron}, A., {Cibirka}, N., {Zitrin}, A., {et~al.} 2018, \apj, 858, 42, \dodoi{10.3847/1538-4357/aabe29}

\bibitem[{{Adam} {et~al.}(2021){Adam}, {Goksu}, {Brown}, {Rudnick}, \& {Ferrari}}]{2021A&A...648A..60A}
{Adam}, R., {Goksu}, H., {Brown}, S., {Rudnick}, L., \& {Ferrari}, C. 2021, \aap, 648, A60, \dodoi{10.1051/0004-6361/202039660}

\bibitem[{{Astropy Collaboration} {et~al.}(2013){Astropy Collaboration}, {Robitaille}, {Tollerud}, {Greenfield}, {Droettboom}, {Bray}, {Aldcroft}, {Davis}, {Ginsburg}, {Price-Whelan}, {Kerzendorf}, {Conley}, {Crighton}, {Barbary}, {Muna}, {Ferguson}, {Grollier}, {Parikh}, {Nair}, {Unther}, {Deil}, {Woillez}, {Conseil}, {Kramer}, {Turner}, {Singer}, {Fox}, {Weaver}, {Zabalza}, {Edwards}, {Azalee Bostroem}, {Burke}, {Casey}, {Crawford}, {Dencheva}, {Ely}, {Jenness}, {Labrie}, {Lim}, {Pierfederici}, {Pontzen}, {Ptak}, {Refsdal}, {Servillat}, \& {Streicher}}]{2013A&A...558A..33A}
{Astropy Collaboration}, {Robitaille}, T.~P., {Tollerud}, E.~J., {et~al.} 2013, \aap, 558, A33, \dodoi{10.1051/0004-6361/201322068}

\bibitem[{{Astropy Collaboration} {et~al.}(2018){Astropy Collaboration}, {Price-Whelan}, {Sip{\H{o}}cz}, {G{\"u}nther}, {Lim}, {Crawford}, {Conseil}, {Shupe}, {Craig}, {Dencheva}, {Ginsburg}, {VanderPlas}, {Bradley}, {P{\'e}rez-Su{\'a}rez}, {de Val-Borro}, {Aldcroft}, {Cruz}, {Robitaille}, {Tollerud}, {Ardelean}, {Babej}, {Bach}, {Bachetti}, {Bakanov}, {Bamford}, {Barentsen}, {Barmby}, {Baumbach}, {Berry}, {Biscani}, {Boquien}, {Bostroem}, {Bouma}, {Brammer}, {Bray}, {Breytenbach}, {Buddelmeijer}, {Burke}, {Calderone}, {Cano Rodr{\'\i}guez}, {Cara}, {Cardoso}, {Cheedella}, {Copin}, {Corrales}, {Crichton}, {D'Avella}, {Deil}, {Depagne}, {Dietrich}, {Donath}, {Droettboom}, {Earl}, {Erben}, {Fabbro}, {Ferreira}, {Finethy}, {Fox}, {Garrison}, {Gibbons}, {Goldstein}, {Gommers}, {Greco}, {Greenfield}, {Groener}, {Grollier}, {Hagen}, {Hirst}, {Homeier}, {Horton}, {Hosseinzadeh}, {Hu}, {Hunkeler}, {Ivezi{\'c}}, {Jain}, {Jenness}, {Kanarek}, {Kendrew}, {Kern}, {Kerzendorf}, {Khvalko}, {King}, {Kirkby}, {Kulkarni},
  {Kumar}, {Lee}, {Lenz}, {Littlefair}, {Ma}, {Macleod}, {Mastropietro}, {McCully}, {Montagnac}, {Morris}, {Mueller}, {Mumford}, {Muna}, {Murphy}, {Nelson}, {Nguyen}, {Ninan}, {N{\"o}the}, {Ogaz}, {Oh}, {Parejko}, {Parley}, {Pascual}, {Patil}, {Patil}, {Plunkett}, {Prochaska}, {Rastogi}, {Reddy Janga}, {Sabater}, {Sakurikar}, {Seifert}, {Sherbert}, {Sherwood-Taylor}, {Shih}, {Sick}, {Silbiger}, {Singanamalla}, {Singer}, {Sladen}, {Sooley}, {Sornarajah}, {Streicher}, {Teuben}, {Thomas}, {Tremblay}, {Turner}, {Terr{\'o}n}, {van Kerkwijk}, {de la Vega}, {Watkins}, {Weaver}, {Whitmore}, {Woillez}, {Zabalza}, \& {Astropy Contributors}}]{2018AJ....156..123A}
{Astropy Collaboration}, {Price-Whelan}, A.~M., {Sip{\H{o}}cz}, B.~M., {et~al.} 2018, \aj, 156, 123, \dodoi{10.3847/1538-3881/aabc4f}

\bibitem[{{Balboni} {et~al.}(2024){Balboni}, {Gastaldello}, {Bonafede}, {Botteon}, {Bartalucci}, {Bourdin}, {Brunetti}, {Cassano}, {De Grandi}, {De Luca}, {Ettori}, {Ghizzardi}, {Gitti}, {Iqbal}, {Johnston-Hollitt}, {Lovisari}, {Mazzotta}, {Molendi}, {Pointecouteau}, {Pratt}, {Riva}, {Rossetti}, {Rottgering}, {Sereno}, {van Weeren}, {Venturi}, \& {Veronesi}}]{2024arXiv240218654B}
{Balboni}, M., {Gastaldello}, F., {Bonafede}, A., {et~al.} 2024, arXiv e-prints, arXiv:2402.18654, \dodoi{10.48550/arXiv.2402.18654}

\bibitem[{{Beresnyak} {et~al.}(2013){Beresnyak}, {Xu}, {Li}, \& {Schlickeiser}}]{2013ApJ...771..131B}
{Beresnyak}, A., {Xu}, H., {Li}, H., \& {Schlickeiser}, R. 2013, \apj, 771, 131, \dodoi{10.1088/0004-637X/771/2/131}

\bibitem[{{Blasi} \& {Colafrancesco}(1999)}]{1999NuPhS..70..495B}
{Blasi}, P., \& {Colafrancesco}, S. 1999, Nuclear Physics B Proceedings Supplements, 70, 495, \dodoi{10.1016/S0920-5632(98)00481-2}

\bibitem[{{Bonafede} {et~al.}(2012){Bonafede}, {Br{\"u}ggen}, {van Weeren}, {Vazza}, {Giovannini}, {Ebeling}, {Edge}, {Hoeft}, \& {Klein}}]{2012MNRAS.426...40B}
{Bonafede}, A., {Br{\"u}ggen}, M., {van Weeren}, R., {et~al.} 2012, \mnras, 426, 40, \dodoi{10.1111/j.1365-2966.2012.21570.x}

\bibitem[{Bonafede {et~al.}(2012)Bonafede, Brüggen, van Weeren, Vazza, Giovannini, Ebeling, Edge, Hoeft, \& Klein}]{10.1111/j.1365-2966.2012.21570.x}
Bonafede, A., Brüggen, M., van Weeren, R., {et~al.} 2012, Monthly Notices of the Royal Astronomical Society, 426, 40, \dodoi{10.1111/j.1365-2966.2012.21570.x}

\bibitem[{{Bonafede} {et~al.}(2018){Bonafede}, {Br{\"u}ggen}, {Rafferty}, {Zhuravleva}, {Riseley}, {van Weeren}, {Farnes}, {Vazza}, {Savini}, {Wilber}, {Botteon}, {Brunetti}, {Cassano}, {Ferrari}, {de Gasperin}, {Orr{\'u}}, {Pizzo}, {R{\"o}ttgering}, \& {Shimwell}}]{2018MNRAS.478.2927B}
{Bonafede}, A., {Br{\"u}ggen}, M., {Rafferty}, D., {et~al.} 2018, \mnras, 478, 2927, \dodoi{10.1093/mnras/sty1121}

\bibitem[{{Bonafede} {et~al.}(2022){Bonafede}, {Brunetti}, {Rudnick}, {Vazza}, {Bourdin}, {Giovannini}, {Shimwell}, {Zhang}, {Mazzotta}, {Simionescu}, {Biava}, {Bonnassieux}, {Brienza}, {Br{\"u}ggen}, {Rajpurohit}, {Riseley}, {Stuardi}, {Feretti}, {Tasse}, {Botteon}, {Carretti}, {Cassano}, {Cuciti}, {Gasperin}, {Gastaldello}, {Rossetti}, {Rottgering}, {Venturi}, \& {Weeren}}]{2022ApJ...933..218B}
{Bonafede}, A., {Brunetti}, G., {Rudnick}, L., {et~al.} 2022, \apj, 933, 218, \dodoi{10.3847/1538-4357/ac721d}

\bibitem[{{Botteon} {et~al.}(2020){Botteon}, {Brunetti}, {van Weeren}, {Shimwell}, {Pizzo}, {Cassano}, {Iacobelli}, {Gastaldello}, {B{\^\i}rzan}, {Bonafede}, {Br{\"u}ggen}, {Cuciti}, {Dallacasa}, {de Gasperin}, {Di Gennaro}, {Drabent}, {Hardcastle}, {Hoeft}, {Mandal}, {R{\"o}ttgering}, \& {Simionescu}}]{2020ApJ...897...93B}
{Botteon}, A., {Brunetti}, G., {van Weeren}, R.~J., {et~al.} 2020, \apj, 897, 93, \dodoi{10.3847/1538-4357/ab9a2f}

\bibitem[{{Botteon} {et~al.}(2022){Botteon}, {Shimwell}, {Cassano}, {Cuciti}, {Zhang}, {Bruno}, {Camillini}, {Natale}, {Jones}, {Gastaldello}, {Simionescu}, {Rossetti}, {Akamatsu}, {van Weeren}, {Brunetti}, {Br{\"u}ggen}, {Groeneveld}, {Hoang}, {Hardcastle}, {Ignesti}, {Di Gennaro}, {Bonafede}, {Drabent}, {R{\"o}ttgering}, {Hoeft}, \& {de Gasperin}}]{2022A&A...660A..78B}
{Botteon}, A., {Shimwell}, T.~W., {Cassano}, R., {et~al.} 2022, \aap, 660, A78, \dodoi{10.1051/0004-6361/202143020}

\bibitem[{{Boxelaar} {et~al.}(2021){Boxelaar}, {van Weeren}, \& {Botteon}}]{2021A&C....3500464B}
{Boxelaar}, J.~M., {van Weeren}, R.~J., \& {Botteon}, A. 2021, Astronomy and Computing, 35, 100464, \dodoi{10.1016/j.ascom.2021.100464}

\bibitem[{{Briggs}(1995)}]{1995PhDT.......238B}
{Briggs}, D.~S. 1995, PhD thesis, New Mexico Institute of Mining and Technology

\bibitem[{{Brown} \& {Rudnick}(2011)}]{2011MNRAS.412....2B}
{Brown}, S., \& {Rudnick}, L. 2011, \mnras, 412, 2, \dodoi{10.1111/j.1365-2966.2010.17738.x}

\bibitem[{{Brunetti}(2017)}]{2017AIPC.1792b0009B}
{Brunetti}, G. 2017, in American Institute of Physics Conference Series, Vol. 1792, 6th International Symposium on High Energy Gamma-Ray Astronomy, 020009, \dodoi{10.1063/1.4968894}

\bibitem[{{Brunetti} \& {Jones}(2014)}]{2014IJMPD..2330007B}
{Brunetti}, G., \& {Jones}, T.~W. 2014, International Journal of Modern Physics D, 23, 1430007, \dodoi{10.1142/S0218271814300079}

\bibitem[{{Brunetti} \& {Lazarian}(2007)}]{2007MNRAS.378..245B}
{Brunetti}, G., \& {Lazarian}, A. 2007, \mnras, 378, 245, \dodoi{10.1111/j.1365-2966.2007.11771.x}

\bibitem[{{Brunetti} \& {Lazarian}(2011)}]{2011MNRAS.410..127B}
---. 2011, \mnras, 410, 127, \dodoi{10.1111/j.1365-2966.2010.17457.x}

\bibitem[{{Brunetti} \& {Lazarian}(2016)}]{2016MNRAS.458.2584B}
---. 2016, \mnras, 458, 2584, \dodoi{10.1093/mnras/stw496}

\bibitem[{{Brunetti} {et~al.}(2001){Brunetti}, {Setti}, {Feretti}, \& {Giovannini}}]{2001MNRAS.320..365B}
{Brunetti}, G., {Setti}, G., {Feretti}, L., \& {Giovannini}, G. 2001, \mnras, 320, 365, \dodoi{10.1046/j.1365-8711.2001.03978.x}

\bibitem[{{Brunetti} {et~al.}(2008){Brunetti}, {Giacintucci}, {Cassano}, {Lane}, {Dallacasa}, {Venturi}, {Kassim}, {Setti}, {Cotton}, \& {Markevitch}}]{2008Natur.455..944B}
{Brunetti}, G., {Giacintucci}, S., {Cassano}, R., {et~al.} 2008, \nat, 455, 944, \dodoi{10.1038/nature07379}

\bibitem[{{Bruno} {et~al.}(2021){Bruno}, {Rajpurohit}, {Brunetti}, {Gastaldello}, {Botteon}, {Ignesti}, {Bonafede}, {Dallacasa}, {Cassano}, {van Weeren}, {Cuciti}, {Di Gennaro}, {Shimwell}, \& {Br{\"u}ggen}}]{2021A&A...650A..44B}
{Bruno}, L., {Rajpurohit}, K., {Brunetti}, G., {et~al.} 2021, \aap, 650, A44, \dodoi{10.1051/0004-6361/202039877}

\bibitem[{{Bruno} {et~al.}(2023){Bruno}, {Botteon}, {Shimwell}, {Cuciti}, {de Gasperin}, {Brunetti}, {Dallacasa}, {Gastaldello}, {Rossetti}, {van Weeren}, {Venturi}, {Russo}, {Taffoni}, {Cassano}, {Biava}, {Lusetti}, {Bonafede}, {Ghizzardi}, \& {De Grandi}}]{2023A&A...678A.133B}
{Bruno}, L., {Botteon}, A., {Shimwell}, T., {et~al.} 2023, \aap, 678, A133, \dodoi{10.1051/0004-6361/202347245}

\bibitem[{{Buch} {et~al.}(2023){Buch}, {Kale}, {Muley}, {Kudale}, \& {Ajithkumar}}]{2023JApA...44...37B}
{Buch}, K.~D., {Kale}, R., {Muley}, M., {Kudale}, S., \& {Ajithkumar}, B. 2023, Journal of Astrophysics and Astronomy, 44, 37, \dodoi{10.1007/s12036-023-09919-x}

\bibitem[{{Buch} {et~al.}(2022){Buch}, {Kale}, {Naik}, {Aragade}, {Muley}, {Kudale}, \& {Ajith Kumar}}]{2022JAI....1150008B}
{Buch}, K.~D., {Kale}, R., {Naik}, K.~D., {et~al.} 2022, Journal of Astronomical Instrumentation, 11, 2250008, \dodoi{10.1142/S2251171722500088}

\bibitem[{{Buote}(2001)}]{2001ApJ...553L..15B}
{Buote}, D.~A. 2001, \apjl, 553, L15, \dodoi{10.1086/320500}

\bibitem[{{Cassano} {et~al.}(2006){Cassano}, {Brunetti}, \& {Setti}}]{2006AN....327..557C}
{Cassano}, R., {Brunetti}, G., \& {Setti}, G. 2006, Astronomische Nachrichten, 327, 557, \dodoi{10.1002/asna.200610587}

\bibitem[{{Cassano} {et~al.}(2010){Cassano}, {Ettori}, {Giacintucci}, {Brunetti}, {Markevitch}, {Venturi}, \& {Gitti}}]{2010ApJ...721L..82C}
{Cassano}, R., {Ettori}, S., {Giacintucci}, S., {et~al.} 2010, \apjl, 721, L82, \dodoi{10.1088/2041-8205/721/2/L82}

\bibitem[{{Cassano} {et~al.}(2013){Cassano}, {Ettori}, {Brunetti}, {Giacintucci}, {Pratt}, {Venturi}, {Kale}, {Dolag}, \& {Markevitch}}]{2013ApJ...777..141C}
{Cassano}, R., {Ettori}, S., {Brunetti}, G., {et~al.} 2013, \apj, 777, 141, \dodoi{10.1088/0004-637X/777/2/141}

\bibitem[{{Cassano} {et~al.}(2023){Cassano}, {Cuciti}, {Brunetti}, {Botteon}, {Rossetti}, {Bruno}, {Simionescu}, {Gastaldello}, {van Weeren}, {Br{\"u}ggen}, {Dallacasa}, {Zhang}, {Akamatsu}, {Bonafede}, {Di Gennaro}, {Shimwell}, {de Gasperin}, {R{\"o}ttgering}, \& {Jones}}]{2023A&A...672A..43C}
{Cassano}, R., {Cuciti}, V., {Brunetti}, G., {et~al.} 2023, \aap, 672, A43, \dodoi{10.1051/0004-6361/202244876}

\bibitem[{{Cavaliere} \& {Fusco-Femiano}(1976)}]{1976A&A....49..137C}
{Cavaliere}, A., \& {Fusco-Femiano}, R. 1976, \aap, 49, 137

\bibitem[{{Chandra} \& {Kanekar}(2017)}]{2017ApJ...846..111C}
{Chandra}, P., \& {Kanekar}, N. 2017, \apj, 846, 111, \dodoi{10.3847/1538-4357/aa85a2}

\bibitem[{{Cornwell} {et~al.}(2008){Cornwell}, {Golap}, \& {Bhatnagar}}]{2008ISTSP...2..647C}
{Cornwell}, T.~J., {Golap}, K., \& {Bhatnagar}, S. 2008, IEEE Journal of Selected Topics in Signal Processing, 2, 647, \dodoi{10.1109/JSTSP.2008.2005290}

\bibitem[{{Cova} {et~al.}(2019){Cova}, {Gastaldello}, {Wik}, {Boschin}, {Botteon}, {Brunetti}, {Buote}, {De Grandi}, {Eckert}, {Ettori}, {Feretti}, {Gaspari}, {Ghizzardi}, {Giovannini}, {Girardi}, {Govoni}, {Molendi}, {Murgia}, {Rossetti}, \& {Vacca}}]{2019A&A...628A..83C}
{Cova}, F., {Gastaldello}, F., {Wik}, D.~R., {et~al.} 2019, \aap, 628, A83, \dodoi{10.1051/0004-6361/201834644}

\bibitem[{{Cuciti} {et~al.}(2018){Cuciti}, {Brunetti}, {van Weeren}, {Bonafede}, {Dallacasa}, {Cassano}, {Venturi}, \& {Kale}}]{2018A&A...609A..61C}
{Cuciti}, V., {Brunetti}, G., {van Weeren}, R., {et~al.} 2018, \aap, 609, A61, \dodoi{10.1051/0004-6361/201731174}

\bibitem[{{Cuciti} {et~al.}(2021){Cuciti}, {Cassano}, {Brunetti}, {Dallacasa}, {de Gasperin}, {Ettori}, {Giacintucci}, {Kale}, {Pratt}, {van Weeren}, \& {Venturi}}]{2021A&A...647A..51C}
{Cuciti}, V., {Cassano}, R., {Brunetti}, G., {et~al.} 2021, \aap, 647, A51, \dodoi{10.1051/0004-6361/202039208}

\bibitem[{{Dallacasa} {et~al.}(2009){Dallacasa}, {Brunetti}, {Giacintucci}, {Cassano}, {Venturi}, {Macario}, {Kassim}, {Lane}, \& {Setti}}]{2009ApJ...699.1288D}
{Dallacasa}, D., {Brunetti}, G., {Giacintucci}, S., {et~al.} 2009, \apj, 699, 1288, \dodoi{10.1088/0004-637X/699/2/1288}

\bibitem[{{de Gasperin} {et~al.}(2017){de Gasperin}, {Intema}, {Shimwell}, {Brunetti}, {Br{\"u}ggen}, {En{\ss}lin}, {van Weeren}, {Bonafede}, \& {R{\"o}ttgering}}]{2017SciA....3E1634D}
{de Gasperin}, F., {Intema}, H.~T., {Shimwell}, T.~W., {et~al.} 2017, Science Advances, 3, e1701634, \dodoi{10.1126/sciadv.1701634}

\bibitem[{{Di Gennaro} {et~al.}(2021){Di Gennaro}, {van Weeren}, {Brunetti}, {Cassano}, {Br{\"u}ggen}, {Hoeft}, {Shimwell}, {R{\"o}ttgering}, {Bonafede}, {Botteon}, {Cuciti}, {Dallacasa}, {de Gasperin}, {Dom{\'\i}nguez-Fern{\'a}ndez}, {En{\ss}lin}, {Gastaldello}, {Mandal}, {Rossetti}, \& {Simionescu}}]{2021NatAs...5..268D}
{Di Gennaro}, G., {van Weeren}, R.~J., {Brunetti}, G., {et~al.} 2021, Nature Astronomy, 5, 268, \dodoi{10.1038/s41550-020-01244-5}

\bibitem[{{Duchesne} {et~al.}(2021){Duchesne}, {Johnston-Hollitt}, \& {Wilber}}]{2021PASA...38...31D}
{Duchesne}, S.~W., {Johnston-Hollitt}, M., \& {Wilber}, A.~G. 2021, \pasa, 38, e031, \dodoi{10.1017/pasa.2021.24}

\bibitem[{{Duchesne} {et~al.}(2024){Duchesne}, {Botteon}, {Koribalski}, {Loi}, {Rajpurohit}, {Riseley}, {Rudnick}, {Vernstrom}, {Andernach}, {Hopkins}, {Kapinska}, {Norris}, \& {Zafar}}]{2024PASA...41...26D}
{Duchesne}, S.~W., {Botteon}, A., {Koribalski}, B.~S., {et~al.} 2024, \pasa, 41, e026, \dodoi{10.1017/pasa.2024.10}

\bibitem[{{Eckert} {et~al.}(2017){Eckert}, {Gaspari}, {Vazza}, {Gastaldello}, {Tramacere}, {Zimmer}, {Ettori}, \& {Paltani}}]{2017ApJ...843L..29E}
{Eckert}, D., {Gaspari}, M., {Vazza}, F., {et~al.} 2017, \apjl, 843, L29, \dodoi{10.3847/2041-8213/aa7c1a}

\bibitem[{{Feretti} {et~al.}(2012){Feretti}, {Giovannini}, {Govoni}, \& {Murgia}}]{2012A&ARv..20...54F}
{Feretti}, L., {Giovannini}, G., {Govoni}, F., \& {Murgia}, M. 2012, \aapr, 20, 54, \dodoi{10.1007/s00159-012-0054-z}

\bibitem[{{Giacintucci} {et~al.}(2011){Giacintucci}, {Dallacasa}, {Venturi}, {Brunetti}, {Cassano}, {Markevitch}, \& {Athreya}}]{2011A&A...534A..57G}
{Giacintucci}, S., {Dallacasa}, D., {Venturi}, T., {et~al.} 2011, \aap, 534, A57, \dodoi{10.1051/0004-6361/201117820}

\bibitem[{{Giacintucci} {et~al.}(2013){Giacintucci}, {Kale}, {Wik}, {Venturi}, \& {Markevitch}}]{2013ApJ...766...18G}
{Giacintucci}, S., {Kale}, R., {Wik}, D.~R., {Venturi}, T., \& {Markevitch}, M. 2013, \apj, 766, 18, \dodoi{10.1088/0004-637X/766/1/18}

\bibitem[{{Giacintucci} {et~al.}(2017){Giacintucci}, {Markevitch}, {Cassano}, {Venturi}, {Clarke}, \& {Brunetti}}]{2017ApJ...841...71G}
{Giacintucci}, S., {Markevitch}, M., {Cassano}, R., {et~al.} 2017, \apj, 841, 71, \dodoi{10.3847/1538-4357/aa7069}

\bibitem[{{Giacintucci} {et~al.}(2005){Giacintucci}, {Venturi}, {Brunetti}, {Bardelli}, {Dallacasa}, {Ettori}, {Finoguenov}, {Rao}, \& {Zucca}}]{2005A&A...440..867G}
{Giacintucci}, S., {Venturi}, T., {Brunetti}, G., {et~al.} 2005, \aap, 440, 867, \dodoi{10.1051/0004-6361:20053016}

\bibitem[{{Giacintucci} {et~al.}(2008){Giacintucci}, {Venturi}, {Macario}, {Dallacasa}, {Brunetti}, {Markevitch}, {Cassano}, {Bardelli}, \& {Athreya}}]{2008A&A...486..347G}
{Giacintucci}, S., {Venturi}, T., {Macario}, G., {et~al.} 2008, \aap, 486, 347, \dodoi{10.1051/0004-6361:200809459}

\bibitem[{{Govoni} {et~al.}(2001){Govoni}, {En{\ss}lin}, {Feretti}, \& {Giovannini}}]{2001A&A...369..441G}
{Govoni}, F., {En{\ss}lin}, T.~A., {Feretti}, L., \& {Giovannini}, G. 2001, \aap, 369, 441, \dodoi{10.1051/0004-6361:20010115}

\bibitem[{{Hoang} {et~al.}(2019){Hoang}, {Shimwell}, {van Weeren}, {Brunetti}, {R{\"o}ttgering}, {Andrade-Santos}, {Botteon}, {Br{\"u}ggen}, {Cassano}, {Drabent}, {de Gasperin}, {Hoeft}, {Intema}, {Rafferty}, {Shweta}, \& {Stroe}}]{2019A&A...622A..20H}
{Hoang}, D.~N., {Shimwell}, T.~W., {van Weeren}, R.~J., {et~al.} 2019, \aap, 622, A20, \dodoi{10.1051/0004-6361/201833900}

\bibitem[{{Ignesti}(2022)}]{2022NewA...9201732I}
{Ignesti}, A. 2022, \na, 92, 101732, \dodoi{10.1016/j.newast.2021.101732}

\bibitem[{{Ignesti} {et~al.}(2020){Ignesti}, {Brunetti}, {Gitti}, \& {Giacintucci}}]{2020A&A...640A..37I}
{Ignesti}, A., {Brunetti}, G., {Gitti}, M., \& {Giacintucci}, S. 2020, \aap, 640, A37, \dodoi{10.1051/0004-6361/201937207}

\bibitem[{{Ignesti} {et~al.}(2023){Ignesti}, {Vulcani}, {Botteon}, {Poggianti}, {Giunchi}, {Smith}, {Brunetti}, {Roberts}, {van Weeren}, \& {Rajpurohit}}]{2023A&A...675A.118I}
{Ignesti}, A., {Vulcani}, B., {Botteon}, A., {et~al.} 2023, \aap, 675, A118, \dodoi{10.1051/0004-6361/202346517}

\bibitem[{{Intema} {et~al.}(2017){Intema}, {Jagannathan}, {Mooley}, \& {Frail}}]{2017A&A...598A..78I}
{Intema}, H.~T., {Jagannathan}, P., {Mooley}, K.~P., \& {Frail}, D.~A. 2017, \aap, 598, A78, \dodoi{10.1051/0004-6361/201628536}

\bibitem[{Joshi(2021)}]{Joshi_21}
Joshi, D. 2021, Master's thesis, National Center for Radio Astrophysics (NCRA-TIFR).
\newblock \url{http://www.ncra.tifr.res.in/~ruta/files/Deepak-Joshi-MPhil-thesis.pdf}

\bibitem[{{Kale} \& {Ishwara-Chandra}(2021)}]{2021ExA....51...95K}
{Kale}, R., \& {Ishwara-Chandra}, C.~H. 2021, Experimental Astronomy, 51, 95, \dodoi{10.1007/s10686-020-09677-6}

\bibitem[{{Kale} {et~al.}(2013){Kale}, {Venturi}, {Giacintucci}, {Dallacasa}, {Cassano}, {Brunetti}, {Macario}, \& {Athreya}}]{2013A&A...557A..99K}
{Kale}, R., {Venturi}, T., {Giacintucci}, S., {et~al.} 2013, \aap, 557, A99, \dodoi{10.1051/0004-6361/201321515}

\bibitem[{{Kale} {et~al.}(2015){Kale}, {Venturi}, {Giacintucci}, {Dallacasa}, {Cassano}, {Brunetti}, {Cuciti}, {Macario}, \& {Athreya}}]{2015A&A...579A..92K}
---. 2015, \aap, 579, A92, \dodoi{10.1051/0004-6361/201525695}

\bibitem[{{Kale} {et~al.}(2022){Kale}, {Parekh}, {Rahaman}, {Joshi}, {Venturi}, {Kolokythas}, {Chibueze}, {Sikhosana}, {Pillay}, \& {Knowles}}]{2022MNRAS.514.5969K}
{Kale}, R., {Parekh}, V., {Rahaman}, M., {et~al.} 2022, \mnras, 514, 5969, \dodoi{10.1093/mnras/stac1649}

\bibitem[{{Kelly}(2007)}]{2007ApJ...665.1489K}
{Kelly}, B.~C. 2007, \apj, 665, 1489, \dodoi{10.1086/519947}

\bibitem[{{Knowles} {et~al.}(2022){Knowles}, {Cotton}, {Rudnick}, {Camilo}, {Goedhart}, {Deane}, {Ramatsoku}, {Bietenholz}, {Br{\"u}ggen}, {Button}, {Chen}, {Chibueze}, {Clarke}, {de Gasperin}, {Ianjamasimanana}, {J{\'o}zsa}, {Hilton}, {Kesebonye}, {Kolokythas}, {Kraan-Korteweg}, {Lawrie}, {Lochner}, {Loubser}, {Marchegiani}, {Mhlahlo}, {Moodley}, {Murphy}, {Namumba}, {Oozeer}, {Parekh}, {Pillay}, {Passmoor}, {Ramaila}, {Ranchod}, {Retana-Montenegro}, {Sebokolodi}, {Sikhosana}, {Smirnov}, {Thorat}, {Venturi}, {Abbott}, {Adam}, {Adams}, {Aldera}, {Bauermeister}, {Bennett}, {Bode}, {Botha}, {Botha}, {Brederode}, {Buchner}, {Burger}, {Cheetham}, {de Villiers}, {Dikgale-Mahlakoana}, {du Toit}, {Esterhuyse}, {Fadana}, {Fanaroff}, {Fataar}, {Foley}, {Fourie}, {Frank}, {Gamatham}, {Gatsi}, {Geyer}, {Gouws}, {Gumede}, {Heywood}, {Hlakola}, {Hokwana}, {Hoosen}, {Horn}, {Horrell}, {Hugo}, {Isaacson}, {Jonas}, {Jordaan}, {Joubert}, {Julie}, {Kapp}, {Kasper}, {Kenyon}, {Kotz{\'e}}, {Kotze}, {Kriek}, {Kriel}, {Krishnan},
  {Kusel}, {Legodi}, {Lehmensiek}, {Liebenberg}, {Lord}, {Lunsky}, {Madisa}, {Magnus}, {Main}, {Makhaba}, {Makhathini}, {Malan}, {Manley}, {Marais}, {Maree}, {Martens}, {Mauch}, {McAlpine}, {Merry}, {Millenaar}, {Mokone}, {Monama}, {Mphego}, {New}, {Ngcebetsha}, {Ngoasheng}, {Ockards}, {Otto}, {Patel}, {Peens-Hough}, {Perkins}, {Ramanujam}, {Ramudzuli}, {Ratcliffe}, {Renil}, {Robyntjies}, {Rust}, {Salie}, {Sambu}, {Schollar}, {Schwardt}, {Schwartz}, {Serylak}, {Siebrits}, {Sirothia}, {Slabber}, {Sofeya}, {Taljaard}, {Tasse}, {Tiplady}, {Toruvanda}, {Twum}, {van Balla}, {van der Byl}, {van der Merwe}, {van Dyk}, {Van Tonder}, {Van Wyk}, {Venter}, {Venter}, {Welz}, {Williams}, \& {Xaia}}]{2022A&A...657A..56K}
{Knowles}, K., {Cotton}, W.~D., {Rudnick}, L., {et~al.} 2022, \aap, 657, A56, \dodoi{10.1051/0004-6361/202141488}

\bibitem[{{Lovisari} {et~al.}(2017){Lovisari}, {Forman}, {Jones}, {Ettori}, {Andrade-Santos}, {Arnaud}, {D{\'e}mocl{\`e}s}, {Pratt}, {Randall}, \& {Kraft}}]{2017ApJ...846...51L}
{Lovisari}, L., {Forman}, W.~R., {Jones}, C., {et~al.} 2017, \apj, 846, 51, \dodoi{10.3847/1538-4357/aa855f}

\bibitem[{{Macario} {et~al.}(2013){Macario}, {Venturi}, {Intema}, {Dallacasa}, {Brunetti}, {Cassano}, {Giacintucci}, {Ferrari}, {Ishwara-Chandra}, \& {Athreya}}]{2013A&A...551A.141M}
{Macario}, G., {Venturi}, T., {Intema}, H.~T., {et~al.} 2013, \aap, 551, A141, \dodoi{10.1051/0004-6361/201220667}

\bibitem[{{McMullin} {et~al.}(2007){McMullin}, {Waters}, {Schiebel}, {Young}, \& {Golap}}]{2007ASPC..376..127M}
{McMullin}, J.~P., {Waters}, B., {Schiebel}, D., {Young}, W., \& {Golap}, K. 2007, in Astronomical Society of the Pacific Conference Series, Vol. 376, Astronomical Data Analysis Software and Systems XVI, ed. R.~A. {Shaw}, F.~{Hill}, \& D.~J. {Bell}, 127

\bibitem[{{Murgia} {et~al.}(2009){Murgia}, {Govoni}, {Markevitch}, {Feretti}, {Giovannini}, {Taylor}, \& {Carretti}}]{2009A&A...499..679M}
{Murgia}, M., {Govoni}, F., {Markevitch}, M., {et~al.} 2009, \aap, 499, 679, \dodoi{10.1051/0004-6361/200911659}

\bibitem[{{Offringa}(2010)}]{2010ascl.soft10017O}
{Offringa}, A.~R. 2010, {AOFlagger: RFI Software}, Astrophysics Source Code Library, record ascl:1010.017.
\newblock \doeprint{1010.017}

\bibitem[{{Osinga} {et~al.}(2024){Osinga}, {van Weeren}, {Brunetti}, {Adam}, {Rajpurohit}, {Botteon}, {Callingham}, {Cuciti}, {de Gasperin}, {Miley}, {R{\"o}ttgering}, \& {Shimwell}}]{2024arXiv240509384O}
{Osinga}, E., {van Weeren}, R.~J., {Brunetti}, G., {et~al.} 2024, arXiv e-prints, arXiv:2405.09384, \dodoi{10.48550/arXiv.2405.09384}

\bibitem[{{Perley} \& {Butler}(2013)}]{2013ApJS..204...19P}
{Perley}, R.~A., \& {Butler}, B.~J. 2013, \apjs, 204, 19, \dodoi{10.1088/0067-0049/204/2/19}

\bibitem[{{Perley} \& {Butler}(2017)}]{2017ApJS..230....7P}
---. 2017, \apjs, 230, 7, \dodoi{10.3847/1538-4365/aa6df9}

\bibitem[{Petrosian(2001)}]{petrosian2001nonthermal}
Petrosian, V. 2001, \apj, 557, 560

\bibitem[{Pfrommer {et~al.}(2008)Pfrommer, En{\ss}lin, \& Springel}]{pfrommer2008simulating}
Pfrommer, C., En{\ss}lin, T.~A., \& Springel, V. 2008, Monthly Notices of the Royal Astronomical Society, 385, 1211

\bibitem[{{Planck Collaboration} {et~al.}(2011{\natexlab{a}}){Planck Collaboration}, {Ade}, {Aghanim}, {Arnaud}, {Ashdown}, {Aumont}, {Baccigalupi}, {Balbi}, {Banday}, {Barreiro}, {Bartelmann}, {Bartlett}, {Battaner}, {Battye}, {Benabed}, {Beno{\^\i}t}, {Bernard}, {Bersanelli}, {Bhatia}, {Bock}, {Bonaldi}, {Bond}, {Borrill}, {Bouchet}, {Brown}, {Bucher}, {Burigana}, {Cabella}, {Cantalupo}, {Cardoso}, {Carvalho}, {Catalano}, {Cay{\'o}n}, {Challinor}, {Chamballu}, {Chary}, {Chiang}, {Chiang}, {Chon}, {Christensen}, {Churazov}, {Clements}, {Colafrancesco}, {Colombi}, {Couchot}, {Coulais}, {Crill}, {Cuttaia}, {da Silva}, {Dahle}, {Danese}, {Davis}, {de Bernardis}, {de Gasperis}, {de Rosa}, {de Zotti}, {Delabrouille}, {Delouis}, {D{\'e}sert}, {Dickinson}, {Diego}, {Dolag}, {Dole}, {Donzelli}, {Dor{\'e}}, {D{\"o}rl}, {Douspis}, {Dupac}, {Efstathiou}, {Eisenhardt}, {En{\ss}lin}, {Feroz}, {Finelli}, {Flores-Cacho}, {Forni}, {Fosalba}, {Frailis}, {Franceschi}, {Fromenteau}, {Galeotta}, {Ganga}, {G{\'e}nova-Santos},
  {Giard}, {Giardino}, {Giraud-H{\'e}raud}, {Gonz{\'a}lez-Nuevo}, {Gonz{\'a}lez-Riestra}, {G{\'o}rski}, {Grainge}, {Gratton}, {Gregorio}, {Gruppuso}, {Harrison}, {Hein{\"a}m{\"a}ki}, {Henrot-Versill{\'e}}, {Hern{\'a}ndez-Monteagudo}, {Herranz}, {Hildebrandt}, {Hivon}, {Hobson}, {Holmes}, {Hovest}, {Hoyland}, {Huffenberger}, {Hurier}, {Hurley-Walker}, {Jaffe}, {Jones}, {Juvela}, {Keih{\"a}nen}, {Keskitalo}, {Kisner}, {Kneissl}, {Knox}, {Kurki-Suonio}, {Lagache}, {Lamarre}, {Lasenby}, {Laureijs}, {Lawrence}, {Le Jeune}, {Leach}, {Leonardi}, {Li}, {Liddle}, {Lilje}, {Linden-V{\o}rnle}, {L{\'o}pez-Caniego}, {Lubin}, {Mac{\'\i}as-P{\'e}rez}, {MacTavish}, {Maffei}, {Maino}, {Mandolesi}, {Mann}, {Maris}, {Marleau}, {Mart{\'\i}nez-Gonz{\'a}lez}, {Masi}, {Matarrese}, {Matthai}, {Mazzotta}, {Mei}, {Meinhold}, {Melchiorri}, {Melin}, {Mendes}, {Mennella}, {Mitra}, {Miville-Desch{\^e}nes}, {Moneti}, {Montier}, {Morgante}, {Mortlock}, {Munshi}, {Murphy}, {Naselsky}, {Nati}, {Natoli}, {Netterfield}, {N{\o}rgaard-Nielsen},
  {Noviello}, {Novikov}, {Novikov}, {Olamaie}, {Osborne}, {Pajot}, {Pasian}, {Patanchon}, {Pearson}, {Perdereau}, {Perotto}, {Perrotta}, {Piacentini}, {Piat}, {Pierpaoli}, {Piffaretti}, {Plaszczynski}, {Pointecouteau}, {Polenta}, {Ponthieu}, {Poutanen}, {Pratt}, {Pr{\'e}zeau}, {Prunet}, {Puget}, {Rachen}, {Reach}, {Rebolo}, {Reinecke}, {Renault}, {Ricciardi}, {Riller}, {Ristorcelli}, {Rocha}, {Rosset}, {Rubi{\~n}o-Mart{\'\i}n}, {Rusholme}, {Saar}, {Sandri}, {Santos}, {Saunders}, {Savini}, {Schaefer}, {Scott}, {Seiffert}, {Shellard}, {Smoot}, {Stanford}, {Starck}, {Stivoli}, {Stolyarov}, {Stompor}, {Sudiwala}, {Sunyaev}, {Sutton}, {Sygnet}, {Taburet}, {Tauber}, {Terenzi}, {Toffolatti}, {Tomasi}, {Torre}, {Tristram}, {Tuovinen}, {Valenziano}, {Vibert}, {Vielva}, {Villa}, {Vittorio}, {Wade}, {Wandelt}, {Weller}, {White}, {White}, {Yvon}, {Zacchei}, \& {Zonca}}]{2011A&A...536A...8P}
{Planck Collaboration}, {Ade}, P.~A.~R., {Aghanim}, N., {et~al.} 2011{\natexlab{a}}, \aap, 536, A8, \dodoi{10.1051/0004-6361/201116459}

\bibitem[{{Planck Collaboration} {et~al.}(2011{\natexlab{b}}){Planck Collaboration}, {Aghanim}, {Arnaud}, {Ashdown}, {Aumont}, {Baccigalupi}, {Balbi}, {Banday}, {Barreiro}, {Bartelmann}, {Bartlett}, {Battaner}, {Benabed}, {Beno{\^\i}t}, {Bernard}, {Bersanelli}, {Bhatia}, {Bock}, {Bonaldi}, {Bond}, {Borrill}, {Bouchet}, {Brown}, {Bucher}, {Burigana}, {Cabella}, {Cantalupo}, {Cardoso}, {Carvalho}, {Catalano}, {Cay{\'o}n}, {Challinor}, {Chamballu}, {Chiang}, {Chon}, {Christensen}, {Churazov}, {Clements}, {Colafrancesco}, {Colombi}, {Couchot}, {Coulais}, {Crill}, {Cuttaia}, {da Silva}, {Dahle}, {Danese}, {de Bernardis}, {de Gasperis}, {de Rosa}, {de Zotti}, {Delabrouille}, {Delouis}, {D{\'e}sert}, {Diego}, {Dolag}, {Donzelli}, {Dor{\'e}}, {D{\"o}rl}, {Douspis}, {Dupac}, {Efstathiou}, {En{\ss}lin}, {Finelli}, {Flores-Cacho}, {Forni}, {Frailis}, {Franceschi}, {Fromenteau}, {Galeotta}, {Ganga}, {G{\'e}nova-Santos}, {Giard}, {Giardino}, {Giraud-H{\'e}raud}, {Gonz{\'a}lez-Nuevo}, {Gonz{\'a}lez-Riestra}, {G{\'o}rski},
  {Gratton}, {Gregorio}, {Gruppuso}, {Harrison}, {Hein{\"a}m{\"a}ki}, {Henrot-Versill{\'e}}, {Hern{\'a}ndez-Monteagudo}, {Herranz}, {Hildebrandt}, {Hivon}, {Hobson}, {Holmes}, {Hovest}, {Hoyland}, {Huffenberger}, {Hurier}, {Jaffe}, {Juvela}, {Keih{\"a}nen}, {Keskitalo}, {Kisner}, {Kneissl}, {Knox}, {Kurki-Suonio}, {Lagache}, {Lamarre}, {Lasenby}, {Laureijs}, {Lawrence}, {Le Jeune}, {Leach}, {Leonardi}, {Liddle}, {Linden-V{\o}rnle}, {L{\'o}pez-Caniego}, {Lubin}, {Mac{\'\i}as-P{\'e}rez}, {Maffei}, {Maino}, {Mandolesi}, {Mann}, {Maris}, {Marleau}, {Mart{\'\i}nez-Gonz{\'a}lez}, {Masi}, {Matarrese}, {Matthai}, {Mazzotta}, {Melchiorri}, {Melin}, {Mendes}, {Mennella}, {Mitra}, {Miville-Desch{\^e}nes}, {Moneti}, {Montier}, {Morgante}, {Mortlock}, {Munshi}, {Murphy}, {Naselsky}, {Natoli}, {Netterfield}, {N{\o}rgaard-Nielsen}, {Noviello}, {Novikov}, {Novikov}, {Osborne}, {Pajot}, {Pasian}, {Patanchon}, {Perdereau}, {Perotto}, {Perrotta}, {Piacentini}, {Piat}, {Pierpaoli}, {Piffaretti}, {Plaszczynski}, {Pointecouteau},
  {Polenta}, {Ponthieu}, {Poutanen}, {Pratt}, {Pr{\'e}zeau}, {Prunet}, {Puget}, {Rebolo}, {Reinecke}, {Renault}, {Ricciardi}, {Riller}, {Ristorcelli}, {Rocha}, {Rosset}, {Rubi{\~n}o-Mart{\'\i}n}, {Rusholme}, {Saar}, {Sandri}, {Santos}, {Schaefer}, {Scott}, {Seiffert}, {Smoot}, {Starck}, {Stivoli}, {Stolyarov}, {Sunyaev}, {Sygnet}, {Tauber}, {Terenzi}, {Toffolatti}, {Tomasi}, {Torre}, {Tristram}, {Tuovinen}, {Valenziano}, {Vibert}, {Vielva}, {Villa}, {Vittorio}, {Wandelt}, {White}, {Yvon}, {Zacchei}, \& {Zonca}}]{2011A&A...536A...9P}
{Planck Collaboration}, {Aghanim}, N., {Arnaud}, M., {et~al.} 2011{\natexlab{b}}, \aap, 536, A9, \dodoi{10.1051/0004-6361/201116460}

\bibitem[{{Rahaman} {et~al.}(2021){Rahaman}, {Raja}, {Datta}, {Burns}, {Alden}, \& {Rapetti}}]{2021MNRAS.505..480R}
{Rahaman}, M., {Raja}, R., {Datta}, A., {et~al.} 2021, \mnras, 505, 480, \dodoi{10.1093/mnras/stab1225}

\bibitem[{{Rajpurohit} {et~al.}(2021{\natexlab{a}}){Rajpurohit}, {Vazza}, {van Weeren}, {Hoeft}, {Brienza}, {Bonnassieux}, {Riseley}, {Brunetti}, {Bonafede}, {Br{\"u}ggen}, {Formann}, {Rajpurohit}, {R{\"o}ttgering}, {Drabent}, {Dom{\'\i}nguez-Fern{\'a}ndez}, {Wittor}, \& {Andrade-Santos}}]{2021A&A...654A..41R}
{Rajpurohit}, K., {Vazza}, F., {van Weeren}, R.~J., {et~al.} 2021{\natexlab{a}}, \aap, 654, A41, \dodoi{10.1051/0004-6361/202141060}

\bibitem[{{Rajpurohit} {et~al.}(2021{\natexlab{b}}){Rajpurohit}, {Brunetti}, {Bonafede}, {van Weeren}, {Botteon}, {Vazza}, {Hoeft}, {Riseley}, {Bonnassieux}, {Brienza}, {Forman}, {R{\"o}ttgering}, {Rajpurohit}, {Locatelli}, {Shimwell}, {Cassano}, {Di Gennaro}, {Br{\"u}ggen}, {Wittor}, {Drabent}, \& {Ignesti}}]{2021A&A...646A.135R}
{Rajpurohit}, K., {Brunetti}, G., {Bonafede}, A., {et~al.} 2021{\natexlab{b}}, \aap, 646, A135, \dodoi{10.1051/0004-6361/202039591}

\bibitem[{{Rajpurohit} {et~al.}(2023){Rajpurohit}, {Osinga}, {Brienza}, {Botteon}, {Brunetti}, {Forman}, {Riseley}, {Vazza}, {Bonafede}, {van Weeren}, {Br{\"u}ggen}, {Rajpurohit}, {Drabent}, {Dallacasa}, {Rossetti}, {Rajpurohit}, {Hoeft}, {Bonnassieux}, {Cassano}, \& {Miley}}]{2023A&A...669A...1R}
{Rajpurohit}, K., {Osinga}, E., {Brienza}, M., {et~al.} 2023, \aap, 669, A1, \dodoi{10.1051/0004-6361/202244925}

\bibitem[{{Rau} \& {Cornwell}(2011)}]{2011A&A...532A..71R}
{Rau}, U., \& {Cornwell}, T.~J. 2011, \aap, 532, A71, \dodoi{10.1051/0004-6361/201117104}

\bibitem[{{Riseley} {et~al.}(2024){Riseley}, {Bonafede}, {Bruno}, {Botteon}, {Rossetti}, {Biava}, {Bonnassieux}, {Loi}, {Vernstrom}, \& {Balboni}}]{2024arXiv240300414R}
{Riseley}, C.~J., {Bonafede}, A., {Bruno}, L., {et~al.} 2024, arXiv e-prints, arXiv:2403.00414, \dodoi{10.48550/arXiv.2403.00414}

\bibitem[{{Sandhu} {et~al.}(2019){Sandhu}, {Raja}, {Rahaman}, {Malu}, \& {Datta}}]{2019JApA...40...17S}
{Sandhu}, P., {Raja}, R., {Rahaman}, M., {Malu}, S., \& {Datta}, A. 2019, Journal of Astrophysics and Astronomy, 40, 17, \dodoi{10.1007/s12036-019-9585-2}

\bibitem[{{Santra} {et~al.}(2024){Santra}, {Kale}, {Giacintucci}, {Markevitch}, {De Luca}, {Bourdin}, {Venturi}, {Dallacasa}, {Cassano}, {Brunetti}, \& {Buch}}]{2024ApJ...962...40S}
{Santra}, R., {Kale}, R., {Giacintucci}, S., {et~al.} 2024, \apj, 962, 40, \dodoi{10.3847/1538-4357/ad1190}

\bibitem[{Sarazin(1986)}]{RevModPhys.58.1}
Sarazin, C.~L. 1986, Rev. Mod. Phys., 58, 1, \dodoi{10.1103/RevModPhys.58.1}

\bibitem[{{Scaife} \& {Heald}(2012)}]{2012MNRAS.423L..30S}
{Scaife}, A. M.~M., \& {Heald}, G.~H. 2012, \mnras, 423, L30, \dodoi{10.1111/j.1745-3933.2012.01251.x}

\bibitem[{{Shimwell} {et~al.}(2014){Shimwell}, {Brown}, {Feain}, {Feretti}, {Gaensler}, \& {Lage}}]{2014MNRAS.440.2901S}
{Shimwell}, T.~W., {Brown}, S., {Feain}, I.~J., {et~al.} 2014, \mnras, 440, 2901, \dodoi{10.1093/mnras/stu467}

\bibitem[{{Srivastava} \& {Singal}(2020)}]{2020MNRAS.493.3811S}
{Srivastava}, S., \& {Singal}, A.~K. 2020, \mnras, 493, 3811, \dodoi{10.1093/mnras/staa520}

\bibitem[{{Sunyaev} \& {Zeldovich}(1972)}]{1972CoASP...4..173S}
{Sunyaev}, R.~A., \& {Zeldovich}, Y.~B. 1972, Comments on Astrophysics and Space Physics, 4, 173

\bibitem[{{Terni de Gregory} {et~al.}(2017){Terni de Gregory}, {Feretti}, {Giovannini}, {Govoni}, {Murgia}, {Perley}, \& {Vacca}}]{2017A&A...608A..58T}
{Terni de Gregory}, B., {Feretti}, L., {Giovannini}, G., {et~al.} 2017, \aap, 608, A58, \dodoi{10.1051/0004-6361/201730878}

\bibitem[{{van Weeren}(2011)}]{2011PhDT........14V}
{van Weeren}, R.~J. 2011, PhD thesis, Leiden Observatory

\bibitem[{{van Weeren} {et~al.}(2019){van Weeren}, {de Gasperin}, {Akamatsu}, {Br{\"u}ggen}, {Feretti}, {Kang}, {Stroe}, \& {Zandanel}}]{2019SSRv..215...16V}
{van Weeren}, R.~J., {de Gasperin}, F., {Akamatsu}, H., {et~al.} 2019, \ssr, 215, 16, \dodoi{10.1007/s11214-019-0584-z}

\bibitem[{{van Weeren} {et~al.}(2016){van Weeren}, {Brunetti}, {Br{\"u}ggen}, {Andrade-Santos}, {Ogrean}, {Williams}, {R{\"o}ttgering}, {Dawson}, {Forman}, {de Gasperin}, {Hardcastle}, {Jones}, {Miley}, {Rafferty}, {Rudnick}, {Sabater}, {Sarazin}, {Shimwell}, {Bonafede}, {Best}, {B{\^\i}rzan}, {Cassano}, {Chy{\.z}y}, {Croston}, {Dijkema}, {En{\ss}lin}, {Ferrari}, {Heald}, {Hoeft}, {Horellou}, {Jarvis}, {Kraft}, {Mevius}, {Intema}, {Murray}, {Orr{\'u}}, {Pizzo}, {Sridhar}, {Simionescu}, {Stroe}, {van der Tol}, \& {White}}]{2016ApJ...818..204V}
{van Weeren}, R.~J., {Brunetti}, G., {Br{\"u}ggen}, M., {et~al.} 2016, \apj, 818, 204, \dodoi{10.3847/0004-637X/818/2/204}

\bibitem[{{Vazza} \& {Botteon}(2024)}]{2024Galax..12...19V}
{Vazza}, F., \& {Botteon}, A. 2024, Galaxies, 12, 19, \dodoi{10.3390/galaxies12020019}

\bibitem[{{Wilber} {et~al.}(2018){Wilber}, {Br{\"u}ggen}, {Bonafede}, {Savini}, {Shimwell}, {van Weeren}, {Rafferty}, {Mechev}, {Intema}, {Andrade-Santos}, {Clarke}, {Mahony}, {Morganti}, {Prandoni}, {Brunetti}, {R{\"o}ttgering}, {Mandal}, {de Gasperin}, \& {Hoeft}}]{2018MNRAS.473.3536W}
{Wilber}, A., {Br{\"u}ggen}, M., {Bonafede}, A., {et~al.} 2018, \mnras, 473, 3536, \dodoi{10.1093/mnras/stx2568}

\bibitem[{{Xie} {et~al.}(2020){Xie}, {van Weeren}, {Lovisari}, {Andrade-Santos}, {Botteon}, {Br{\"u}ggen}, {Bulbul}, {Churazov}, {Clarke}, {Forman}, {Intema}, {Jones}, {Kraft}, {Lal}, {Mroczkowski}, \& {Zitrin}}]{2020A&A...636A...3X}
{Xie}, C., {van Weeren}, R.~J., {Lovisari}, L., {et~al.} 2020, \aap, 636, A3, \dodoi{10.1051/0004-6361/201936953}

\end{thebibliography}
\bibliographystyle{aasjournal}



\end{document}